\documentclass[12pt]{iopart}
\usepackage{iopams,eucal,amsthm,mathrsfs,stmaryrd,tensor,xcolor,tikz-cd,cite}
\usepackage[symbol]{footmisc}
\makeatletter
\renewcommand{\thefootnote}{{\@arabic\c@footnote}}
\makeatother
\usepackage{hyperref}
\newtheorem{definition}{Definition}
\newtheorem{theorem}{Theorem}
\newcommand{\ie}{\emph{i.e.}}
\newcommand{\eg}{\emph{e.g.}}
\newcommand{\Eqref}[1]{(\ref{#1})}
\newcommand{\RR}{\mathbb{R}}
\newcommand{\Sph}{\mathbf{S}}
\newcommand{\iim}{\Leftrightarrow}
\newcommand{\order}{\boldsymbol{\CMcal{O}}}

\newcommand{\man}{\mathscr{M}}
\newcommand{\scrH}{\CMcal{H}}
\newcommand{\nhood}{\mathscr{N}_{\CMcal{H}}}
\newcommand{\Spi}{\boldsymbol{\mathsf{Spi}}}
\newcommand{\spi}{\mathfrak{spi}}
\newcommand{\Str}{\boldsymbol{\mathsf{Str}}_{(\scrH)}}
\newcommand{\str}{\mathfrak{str}_{(\scrH)}}
\newcommand{\trn}{\mathfrak{trn}_{(\scrH)}}
\newcommand{\so}{\mathfrak{so}}
\newcommand{\Lorentz}{\boldsymbol{\mathsf{Lorentz}}}
\newcommand{\dl}{\partial}
\newcommand{\Lie}{\pounds}
\newcommand{\dd}{\mathrm{d}}
\newcommand{\Dl}{\nabla}

\newcommand{\met}{\mathsf{g}}
\newcommand{\hmet}{\mathsf{h}}
\newcommand{\pmet}{\mathsf{p}}
\newcommand{\qmet}{\mathsf{q}}

\newcommand{\Ric}{R}
\newcommand{\EinG}{G}
\newcommand{\ehat}{\boldsymbol{\hat{\mathsf{e}}}}
\newcommand{\fhat}{\boldsymbol{\hat{\mathsf{f}}}}
\newcommand{\omegaf}{\omega_{\mathsf{f}}}

\newcommand{\muH}{\underline{\mu}}
\newcommand{\rs}{r_{\mathrm{s}}}

\renewcommand{\bTheta}{\boldsymbol{\mathsf{\Theta}}}
\newcommand{\bThetao}{\boldsymbol{\mathsf{\Theta_{o}}}}
\newcommand{\eqv}[1]{(\kern-2pt({#1})\kern-2pt)}
\newcommand{\ripple}[1]{\llbracket{#1}\rrbracket}
\newcommand{\ST}{\CMcal{T}}

\newcommand{\bn}{\boldsymbol{\check{n}}}
\newcommand{\taun}{\tau_{\perp}}
\newcommand{\taup}{\tau_{\shortparallel}}

\newcommand{\Lan}{\Lambda_{\perp}}
\newcommand{\Lap}{\Lambda_{\shortparallel}}

\newcommand{\upsi}{\underline{\psi}}
\newcommand{\scA}{\textsc{a}}
\newcommand{\scI}{\textsc{i}}
\newcommand{\scN}{\textsc{n}}
\newcommand{\calN}{\CMcal{N}}
\newcommand{\pertS}{\boldsymbol{\mathsf{s}}}
\newcommand{\pertT}{\boldsymbol{\mathsf{t}}}
\newcommand{\pertV}{\boldsymbol{\mathsf{v}}}
\newcommand{\KVT}{\mathcal{T}}
\begin{document}

\title[Asymptotic symmetries at spatial infinity]{Asymptotic symmetries at spatial infinity}

\author{Sharad Mishra\footnote{Corresponding author}}
\address{Department of Physics, \\ Birla Institute of Technology and Science, Pilani, \\ K K Birla Goa Campus, \\ Zuarinagar, Sancoale, Goa 403726, India}
\ead{p20200069@goa.bits-pilani.ac.in}
\vspace{10pt}

\author{Kinjal Banerjee}
\address{Department of Physics, \\ Birla Institute of Technology and Science, Pilani, \\ K K Birla Goa Campus, \\ Zuarinagar, Sancoale, Goa 403726, India}
\ead{kinjalb@gmail.com}
\vspace{10pt}

\author{Jishnu Bhattacharyya}
\ead{jishnub@gmail.com}
\begin{abstract}
We describe asymptotic symmetries at spatial infinity of asymptotically flat spacetimes within the context of a generalization of the Beig-Schmidt-Ashtekar-Romano-framework. We demonstrate that it is possible to relax certain smoothness requirements of the asymptotic transformations considered previously, without violating asymptotic flatness. This leads to an enhancement of the asymptotic symmetry group that includes logarithmic supertranslations at spatial infinity. Our results compliment several recent results which confirm the existence of logarithmic supertranslations at spatial infinity in the Hamiltonian formalism.
\end{abstract}
\section{Introduction}
It is reasonable to expect that a spacetime satisfying Einstein's field equations should appear approximately flat `sufficiently far away' from `localized' matter. This intuition is formalized into various definitions of~\emph{asymptotic flatness}. In particular, depending on the causal nature of the separation of the `localized' matter from the `distant observer', the notions of asymptotic flatness near~\emph{timelike}-,~\emph{null}- or~\emph{spatial infinity} are formulated.

Historically, the central motivation behind carrying out the above program near null infinity was to understand the nature of gravitational radiation `far away' from gravitating objects~\cite{Bondi:1962px, Sachs:1961zz, Sachs:1962wk, Sachs:1962zza, Penrose:1962ij, Penrose:1964ge, Penrose:1965am} (a very useful review is~\cite{Ashtekar:2014zsa}). The current -- and vigorously ongoing -- interest in asymptotically flat spacetimes stems from the recently discovered novel connections between asymptotic symmetries at null infinity, soft theorems and the memory effect~\cite{Strominger:2013jfa, Strominger:2013lka, He:2014cra, He:2014laa, Strominger:2014pwa, Pasterski:2015tva, Pasterski:2015zua, Susskind:2015hpa} (the lecture notes~\cite{Raclariu:2021zjz, Pasterski:2021rjz} provide reasonably recent accounts of the story and a fairly detailed guide into the literature).

In contrast, the study of the region near spatial infinity was originally motivated by the need to determine consistent boundary conditions for physical fields on `constant time slices' within the context of initial value formulation of general relativity~\cite{Arnowitt:1959ah, Arnowitt:1960es, Arnowitt:1960zzc, Arnowitt:1961zz, Arnowitt:1961zza, Arnowitt:1962hi, 1962gicr.book..227A, OMurchadha:1974mtn, OMurchadha:1974xtz} (see also~\cite{Regge:1974zd, Beig:1987zz, Ashtekar:1990gc} for related work). Such attempts were geometrized and eventually covariantized in~\cite{Geroch:1972up, Geroch:1977big, Ashtekar:1978zz, paulsommers, Persides:1979sge, Persides:1980jds, Persides:1980otb, 1980grg2.conf...37A, Beig:1982ifu, Beig:1983sw, Ashtekar:1991vb} (for further details, refer to the slightly dated but still very informative and insightful discussions in~\cite{Geroch:1977big, 1980grg2.conf...37A, Ashtekar:1991vb}).

Despite all this progress, a complete understanding of the nature of~\emph{asymptotic symmetries at spatial infinity} is still lacking, as is evidenced from the current research efforts in this direction~\cite{Troessaert:2017jcm, Henneaux:2018cst, Henneaux:2018gfi, Henneaux:2018hdj, Henneaux:2019yax, Fuentealba:2022xsz, Fuentealba:2023syb,solankisrijith,Ali_Mohamed_2021,Chakraborty_2025}. It has long been understood that the group of asymptotic symmetries at spatial infinity include an infinite dimensional subgroup which is very similar to the~\emph{BMS symmetry} group at null infinity~\cite{Ashtekar:1978zz}. However, the physical significance of this symmetry group, and especially the connection between the generators of the corresponding Lie algebra and those of the infinitesimal BMS symmetry transformations at null infinity, remain elusive. Intimately tied to this topic is a satisfactory resolution of the so called `logarithmic ambiguities' at spatial infinity~\cite{PhysRev.124.274, ashtekar1985logarithmic, Chrusciel:1989ye, Ashtekar:2008jw, Ashtekar:1991vb} and their connection to what has been termed as~\emph{logarithmic supertranslations} in the recent literature (see, \eg,~\cite{Compere:2011ve, Troessaert:2017jcm, Fuentealba:2022xsz}).

The present work describes first steps towards addressing some of the aforementioned issues within the context of an extension of the framework introduced in~\cite{Ashtekar:1991vb}. This approach can be viewed as a `covariantization' of the explicit coordinate dependent treatment of~\cite{Beig:1982ifu, Beig:1983sw} (in the same spirit as Penrose's conformal description of null infinity is a `covariantization' of Bondi~\emph{et. al.}'s description via null coordinates). In what follows, we will refer to this framework as the~\emph{Beig-Schmidt-Ashtekar-Romano}- or~\emph{BSAR}-framework. The basic idea behind this framework is very simple:
	\begin{itemize}
	 \item One postulates (see below) that near spatial infinity,~\emph{every} asymptotically flat spacetime admits a foliation which resembles the hyperbolic foliation of global Minkowski spacetime (akin to two-sphere foliation of global three-dimensional Euclidean space).
	 \item The hypersurface described by the limiting value of the corresponding `hyperbolic-radial coordinate function' represents the boundary at spatial infinity\footnote{The boundary itself is not a part of the physical spacetime. The situation is very much like how the `infinitely large' two-sphere can be intuitively thought of as the boundary of three dimensional Euclidean space.}. Hence spatial infinity appears as a genuine boundary (\ie, a codimension one hypersurface) of asymptotically flat spacetimes in this framework.
	 \item A collection of geometric quantities naturally associated with such a foliation, defined with or without the help of the physical metric, is recognized to describe~\emph{the universal structure} associated with spatial infinity of every asymptotically flat spacetime.
	 \item A diffeomorphism preserving this universal structure is regarded as an asymptotic symmetry transformation. The collection of all such transformation forms the asymptotic symmetry group.
	\end{itemize}
Our work is heavily based on reference~\cite{Ashtekar:1991vb} and describes an attempt to extend the BSAR-framework. More specifically, we show that certain `smoothness requirements' previously imposed on the asymptotic symmetry transformations can be consistently relaxed without affecting the essential properties of the asymptotic symmetry group. This results in an enlargement of the group by allowing the existence of the logarithmic supertranslations.

The rest of this paper is organized as follows: in Section~\ref{sec:bsar}, we recall the definitions of asymptotic flatness near spatial infinity following~\cite{Ashtekar:1991vb}. These definitions formalize the heuristic ideas presented earlier (about the existence of foliations resembling the hyperbolic foliation of global Minkowski spacetime) through the concept of~\emph{completions}. We also introduce various~\emph{kinematic} (geometric) quantities associated with a completion and discuss some of their properties.

In Section~\ref{sec:UHC}, we emphasize the non-uniqueness of completions and identify a special class of them -- called~\emph{unit hyperboloid completions} -- for their physical significance. The transformations between such completions, as well as between the kinematic quantities associated with such completions, are studied. In particular, we demonstrate how to systematically relax the `smoothness restrictions' on these transformations (which were previously assumed in~\cite{Ashtekar:1991vb}) and discover that they can become `too general' unless we restrict them to affect a dynamically relevant quantity -- called the~\emph{mass aspect} -- in a particular way. This physical restriction is then shown to lead to the logarithmic supertranslations.

In Section~\ref{sec:Str} we establish that the set of transformations studied in the previous section admits a group structure. We further demonstrate that this group may become non-Abelian under the generalized transformations, but the physical restrictions stemming from the behavior of the mass aspect prevents that outcome. We argue why this indicates consistency of the generalized transformations. We also formulate the criteria for~\emph{gauge transformations} and introduce the notion of~\emph{the universal structure} associated with every asymptotically flat spacetime.

Finally, in Section~\ref{sec:ASAM} we focus on identifying the set of~\emph{all} transformations which leave the universal structure invariant. This leads us to the notion of an~\emph{asymptotic symmetry generator}. We define these generators, establish their essential properties (including how to identify~\emph{supertranslations},~\emph{asymptotic Lorentz generators} and~\emph{gauge generators}), study the form of the Lie algebra that they generate, and argue why our findings can be considered as a consistent generalization of~\cite{Ashtekar:1991vb}.

We conclude this paper in Section~\ref{sec:outlook} by highlighting the ways the current studies can be further extended. We emphasize the kinematical nature of our approach and contrast that with other approaches of studying asymptotic symmetries which focus more on the underlying dynamics of the physical theory.

We have included several appendices for the proofs of various theorems and derivations of some of the important results presented in the main text. In the final appendix, we demonstrate how to construct a unit hyperboloid completion of the Schwarzschild spacetime as well as compute some of the associated kinematical quantities for that specific case.
\section{Spatial infinity as a boundary}\label{sec:bsar}
The conditions for a spacetime to be asymptotically flat/Minkowskian at spatial infinity were precisely formulated in~\cite{Ashtekar:1991vb}. We quote these definitions almost verbatim below with a slight change in notation\footnote{Our notations for the physical metric and its inverse are $\met_{a b}$ and $\met^{a b}$ respectively, whereas~\cite{Ashtekar:1991vb} uses the notations $\hat{\met}_{a b}$ and $\hat{\met}^{a b}$ for them. For us the physical spacetime manifold is $\man$, and the unphysical manifold whose boundary represents the boundary at spatial infinity is $\hat{\man}$; for~\cite{Ashtekar:1991vb} it is the other way around.}.
\begin{definition}[Asymptote at spatial infinity]\label{def:AR:asymp}
A spacetime $(\man, \met)$ will be said to posses~\emph{an asymptote at spatial infinity} if there exists a manifold $\hat{\man}$ with a boundary $\scrH$, a function $\Omega$ defined on $\hat{\man}$, and a diffeomorphism from $\man$ to $\hat{\man} \setminus \scrH$ (with which we identify $\man$ and its image in $\hat{\man}$) such that
	\begin{enumerate}
	 \item\label{def:AR:asymp_i} $\Omega~\hat{=}~0$, $\Dl_{a}\Omega~\hat{\neq}~0$~\emph{(}where the symbol~\emph{`}$\hat{=}$\emph{'} stands for~\emph{`}equals when evaluated at points of $\scrH$\emph{')}, and $\Dl_{a}\Omega$ is at least $C^{3}$ in a suitable neighborhood $\nhood$ of $\scrH$;
	 \item\label{def:AR:asymp_ii} The vector
	  \begin{equation}\label{def:n^a}
	   n^{a}~\equiv~\Omega^{-4}\met^{a b}\Dl_{b}\Omega~,
	  \end{equation}
and the rank-$(0, 2)$ symmetric tensor
	  \begin{equation}\label{def:qmet_ab}
	   \qmet_{a b}~\equiv~\Omega^{2}\left(\met_{a b} - F^{-1}\Omega^{-4}\Dl_{a}\Omega\Dl_{b}\Omega\right)~,
	  \end{equation}
where
	  \begin{equation}\label{def:F}
	   F~\equiv~n^{a}\Dl_{a}\Omega~,
	  \end{equation}
are at least $C^{3}$ in a suitable neighborhood $\nhood$ of $\scrH$ as well as on $\scrH$, and they both admit finite limits to $\scrH$ such that $\qmet_{a b}$ has signature $(-,\,+,\,+)$ on $\scrH$;
	 \item $\displaystyle{\lim_{\Omega \to 0}}\Omega^{-1}\EinG_{a b}~\hat{=}~0$, where $\EinG_{a b} \equiv \Ric_{a b} -\frac{1}{2}\Ric\met_{a b}$ is the Einstein tensor of~\emph{(the physical metric)} $\met_{a b}$.
	\end{enumerate}
\end{definition}
In what follows, the pair $(\hat{\man}, \Omega)$ will be called a~\emph{completion} of the (physical) spacetime $(\man, \met)$, and the corresponding function $\Omega$ will be called a~\emph{completion function}.
\begin{definition}[AFSI spacetime]\label{def:AR:AFSI}
If $(\man, \met)$ admits an asymptote at spatial infinity in which $\scrH$ has topology $\RR \times \Sph^{2}$ then $(\man, \met)$ will be said to be~\emph{asymptotically flat at spatial infinity (AFSI)}.
\end{definition}
\begin{definition}[AM spacetime]\label{def:AR:AM}
An AFSI spacetime $(\man, \met)$ will be said to be~\emph{asymptotically Minkowskian (AM)} if the boundary $\scrH$ is geodesically complete with respect to the metric $\qmet_{a b}$ thereon.
\end{definition}
\paragraph{Remarks:}
	\begin{itemize}
	 \item In this work, the differentiability of various geometrical objects introduced above need not be stronger than ${C}^3$. This is in contrast to the requirements in reference~\cite{Ashtekar:1991vb} who assumed these geometrical objects to be `smooth' (\ie, `differentiable as many times as desired'). The weakening of the smoothness requirements lead to an enlargement of the set of asymptotic symmetry generators, as will become clear below.
	 \item Our calculations are covariant and we will often work with various tensorial objects in some appropriate neighborhood of $\scrH$ characterized by~\emph{sufficiently small values of $\Omega$} in a given completion $(\hat{\man}, \Omega)$. Unless otherwise specified, the notation $\nhood$ (introduced in Definition~\ref{def:AR:asymp}) will always denote such a neighborhood.
\item We will also frequently use the phrase~`\emph{asymptotic limit}' below to denote a limit to (and only to) $\scrH$.
	 \item As~\cite{Ashtekar:1991vb} explains, the (weaker) Definitions~\ref{def:AR:asymp} and~\ref{def:AR:AFSI} (which are, nevertheless, prerequisites for Definition~\ref{def:AR:AM}), provide adequate local structures at infinity (\eg, to arrive at results which do not require the strongest global structure offered by Definition~\ref{def:AR:AM}). However, we will explicitly assume Definition~\ref{def:AR:AM} while discussing asymptotic symmetries. This is because, while the Lie algebra associated with the `infinitesimal' symmetry transformations require only Definition~\ref{def:AR:asymp}, the existence of the corresponding symmetry group depends crucially on the global structure at infinity. However, most of the following discussion, being local in nature, will essentially depend on Definition~\ref{def:AR:asymp}.
	\end{itemize}
Let us now look at some of the consequences of the above definitions in detail (the following observations already appear in reference~\cite{Ashtekar:1991vb}; we recount them here for the sake of a coherent presentation).

Definition~\ref{def:AR:asymp}\Eqref{def:AR:asymp_i} guarantees that the one-form $\Dl_{a}\Omega$ is defined on the $\Omega = 0$ hypersurface (which, by definition, is the boundary $\scrH$) and it is non-vanishing there. Likewise, the purpose of Definition~\ref{def:AR:asymp}\Eqref{def:AR:asymp_ii} is to ensure that the defining expressions for the vector $n^{a}$ and the rank-$(0, 2)$ symmetric tensor $\qmet_{a b}$, which are well-defined in the physical spacetime, also admit finite limits to $\scrH$. Consequently $F$ also admits a finite limit to $\scrH$ (we shall show later that $F$ is non-zero on $\scrH$). The physical metric can now be expressed as
	\begin{equation}\label{expr:met_ab}
	 \met_{a b} = F^{-1}\Omega^{-4}\Dl_{a}\Omega\Dl_{b}\Omega + \Omega^{-2}\qmet_{a b}~,
	\end{equation}
in the bulk of the spacetime (neither side admits a finite limit to $\scrH$). To decompose the inverse physical metric in the same spirit, define
	\begin{equation}\label{def:qmet^ab}
	 \qmet^{a b}~\equiv~\Omega^{-4}\met^{a c}\met^{b d}\qmet_{c d}~,
	\end{equation}
from which it follows
	\begin{equation}\label{expr:met^ab}
	 \met^{a b} = F^{-1}\Omega^{4}n^{a}n^{b} + \Omega^{2}\qmet^{a b}~.
	\end{equation}
We will later argue that  $\qmet^{a b}$ admits a finite limit to $\scrH$ as well. Therefore, $\met^{a b}$ is well defined everywhere in the physical spacetime, and vanishes only on $\scrH$.

The various definitions introduced above imply
	\begin{equation*}
	 \qmet_{a b}n^{b} = 0~, \qquad\qquad\qquad\qquad \qmet^{a b}\Dl_{b}\Omega = 0~,
	\end{equation*}
indicating that the decompositions of the physical metric and its inverse presented above are, in fact, orthogonal decompositions. We may further observe
	\begin{equation*}
	 \Omega^{2}\qmet^{a b} = \met^{a c}\met^{b d}\left(\Omega^{-2}\qmet_{c d}\right) \qquad\qquad\iim\qquad\qquad \Omega^{-2}\qmet_{a b} = \met_{a c}\met_{b d}\left(\Omega^{2}\qmet^{c d}\right),
	\end{equation*}
which shows that the tensor $q_{a b} \equiv \Omega^{-2}\qmet_{a b}$ is essentially the `three-metric' induced on the $\Omega =$ constant ($\neq 0$) hypersurfaces by the physical metric, and $q^{a b} \equiv \Omega^{2}\qmet^{a b}$ is its `inverse'\footnote{Reference~\cite{Ashtekar:1991vb} uses the notations $\hat{q}_{a b}$ and $\hat{q}^{a b}$ for $q_{a b}$ and $q^{a b}$, respectively.}. This becomes even more transparent through the following identities introducing the projector on the $\Omega =$ constant surfaces
	\begin{equation}\label{def:pmet}
	 \tensor{\pmet}{^{a}_{b}}~\equiv~\qmet^{a c}\qmet_{c b} = q^{a c}q_{c b}~, \qquad\qquad  \tensor{\delta}{^{a}_{b}} - F^{-1}n^{a}\Dl_{b}\Omega = \tensor{\pmet}{^{a}_{b}}~.
	\end{equation}
Since $\qmet_{a b}$ admits a finite limit to $\scrH$ by definition, so does $\qmet^{a b}$ as well as the projector.

In the bulk of the spacetime, the (squared) norm of the vector $n^{a}$ is given by 
	\begin{equation*}
	 \met_{a b}n^{a}n^{b} = \Omega^{-4}F~.
	\end{equation*}
Since the left hand side need not admit a finite limit to $\scrH$ (in fact, it doesn't) neither does the right hand side. However, the above expression makes it clear that the sign of the function $F$ determines the causal nature of the vector $n^{a}$ near $\scrH$ -- and hence on $\scrH$, by continuity. The same is true for the one-form $\Dl_{a}\Omega$ based on the relation
	\begin{equation*}
	 \met^{a b}\Dl_{a}\Omega\Dl_{b}\Omega = \Omega^{4}F~.
	\end{equation*}
Now, Definition~\ref{def:AR:AFSI} dictates that $\qmet_{a b}$ has signature $(-,\,+,\,+)$ on $\scrH$. However, since $\qmet_{a b}$ is also $C^{3}$ in $\nhood$, it has signature $(-,\,+,\,+)$ in the same neighborhood as well. At any point in $\nhood$, there must then exist a vector $v^{a}$ satisfying $v^{a}\Dl_{a}\Omega = 0$ (\ie, a vector lying on some $\Omega =$ constant hypersurface) as well as $\qmet_{a b}v^{a}v^{b} < 0$ (due to the assumed signature of $\qmet_{a b}$). Hence $v^{a}$ must be timelike (with respect to the physical metric) at that point. However, since the chosen point is generic, the conclusion should hold everywhere in $\nhood$. Consequently, $F > 0$ in $\nhood$, and therefore also~\emph{on} $\scrH$ by continuity.

Since $q_{a b}$ is essentially the metric induced by the physical metric on the $\Omega =$ constant ($\neq 0$) surfaces, the covariant derivative $D_{a}$ compatible with $q_{a b}$ is simply the projection of the covariant derivative $\Dl_{a}$ compatible with the physical metric onto such $\Omega =$ constant surfaces. Since $D_{a}\Omega = 0$, however, $D_{a}$ is also compatible with $\qmet_{a b}$, \ie
	\begin{equation*}
	 D_{a}\qmet_{b c} = 0~.
	\end{equation*}
The covariant derivative $D_{a}$ extends to $\scrH$ since the projectors admit analogous limits. Looking forward, our construction will depend on $\Dl_{a}\Omega$, $n^{a}$, $\qmet_{a b}$ and $\qmet^{a b}$ all admitting well defined limits to $\scrH$.

There exists a special coordinate system associated with a completion $(\hat{\man}, \Omega)$ which employs the function $\Omega$ as one of coordinates (with the remaining three being some suitable set of functions coordinatizing the $\Omega =$ constant hypersurfaces). Such a coordinate system is specifically adapted to the orthogonal decomposition of the physical metric~\Eqref{expr:met_ab} and is closely related to the~\emph{Beig-Schmidt coordinate system}~\cite{Beig:1982ifu, Ashtekar:1991vb}. The following vector field generates `$\Omega$-translations' in such a coordinate system
	\begin{equation}\label{def:d/dO}
	 \left(\frac{\dl}{\dl\Omega}\right)^{a}~\equiv~F^{-1}n^{a}~.
	\end{equation}
and admits a well defined extension to $\scrH$ by construction. While we will not commit to any coordinate system in what follows, we will nevertheless denote a directional/Lie derivative of a given scalar function along the above vector simply as the `$\Omega$-derivative' of the said scalar.

\section{Unit hyperboloid completions}\label{sec:UHC}
As per Definition~\ref{def:AR:asymp},~\emph{any} choice of the function $\Omega$ satisfying
	\begin{equation*}
	 \Omega~\hat{=}~0~, \qquad\qquad \Dl_{a}\Omega~\hat{\neq}~0~,
	\end{equation*}
is a valid completion; \ie, $\Omega$ is not unique. Consider two such completions of a physical spacetime $(\man, \met)$, given by  $(\hat{\man}, \Omega)$ and $(\hat{\man}, \Omega')$. It should be natural to expect some sort of `democracy' among the completions, in the sense that there should be nothing special about any choice of the completion function. Hence going from any valid completion to another one must be allowed. In particular, this means that if
	\begin{equation}\label{def:Omega'}
	 \Omega'~\equiv~\alpha\Omega~,
	\end{equation}
then
	\begin{equation}\label{def:alpha'}
	 \Omega = \alpha'\Omega'~,
	\end{equation}
such that
	\begin{equation}\label{expr:alpha'alpha}
	 \alpha'\alpha = 1~.     
	\end{equation}
Reference~\cite{Ashtekar:1991vb} assumes $\alpha$ to be a smooth function. We want to relax such smoothness requirements as much as possible while ensuring finiteness of the above kinematical quantities on $\scrH$ in every completion. Now, if $\alpha$ vanishes on $\scrH$, say, then $\alpha'$ will diverge there according to the above equation, and~\emph{vice-versa}. In order to prevent such undesirable behavior, our immediate goal will be to find appropriate conditions such that the functions $\alpha$ and $\alpha'$ both admit finite and non-zero asymptotic limits. Subsequently, we will also derive the transformation rules for the kinematical quantities $\Dl_{a}\Omega$, $n^{a}$, $\qmet_{a b}$ and $\qmet^{a b}$, under~\Eqref{def:Omega'}, given the aforementioned restrictions on $\alpha$ and $\alpha'$.

To that end, since both $\Dl_{a}\Omega$ and $\Dl_{a}\Omega'$ must exist on $\scrH$ by definition, we must require
	\begin{itemize}
	 \item $\alpha$ be at least $C^{1}$ in $\nhood$,
	\end{itemize}
such that
	\begin{equation}\label{dO':gen}
	 \Dl_{a}\Omega' = \beta\Dl_{a}\Omega + \Omega D_{a}\alpha~,
	\end{equation}
where
	\begin{equation}\label{def:beta}
	 \beta~\equiv~\alpha + \Omega\frac{\dl\alpha}{\dl\Omega}~.
	\end{equation}
We will also require that in the asymptotic limit
	\begin{itemize}
	 \item $\Dl_{a}\alpha$ is allowed to diverge asymptotically (so that more general transformations between the completion functions compared to reference~\cite{Ashtekar:1991vb} are allowed); 
	 \item $\Omega\Dl_{a}\alpha$ must vanish.
	\end{itemize}
The final requirement thus ensures that the gradients of the completion functions are parallel to each other everywhere on $\scrH$ such that
	\begin{equation}\label{beta=alpha:H}
	 \beta~\hat{=}~\alpha~, \qquad\qquad\qquad \Dl_{a}\Omega'~\hat{=}~\alpha\Dl_{a}\Omega~.
	\end{equation}
The remaining restrictions on $\alpha$ will follow from demanding finiteness of the kinematic quantities under change in completions  on $\scrH$.

We are going to adopt a notational convention where the kinematic quantities associated with the completion $(\hat{\man}, \Omega')$ will appear with an additional `$\prime$' symbol, but otherwise will have the same definitions as the corresponding ones associated with $(\hat{\man}, \Omega)$. Using the relations derived in Section~\ref{sec:bsar}, it can be easily seen that 
	\begin{equation}\label{n'^a:gen}
	 n^{\prime a} = \alpha^{-4}\left(\beta n^{a} + \Omega^{-1}D^{a}\alpha\right),
	\end{equation}
in $\nhood$, where $D^{a}f \equiv \qmet^{a b}D_{b}f$ for any function $f$ here and henceforth. The right hand side can only admit a finite limit to $\scrH$ only if
	\begin{equation}\label{Dalpha:H}
	 D_{a}\alpha~\hat{=}~0~.
	\end{equation}
Thus, the final restrictions on $\alpha$ required is
	\begin{itemize}
    	 \item $\alpha$ is constant on $\scrH$.
	\end{itemize}
A similar straightforward calculation yields
	\begin{equation}\label{Fprime-alpha}
	 F' = \alpha^{-4}\left[\beta^{2}F + (D^{c}\alpha)(D_{c}\alpha)\right],
	\end{equation}
which, on using~\Eqref{beta=alpha:H} and~\Eqref{Dalpha:H}, gives 
	\begin{equation}\label{F':H}
	 F'~\hat{=}~\alpha^{-2}F~.
	\end{equation}
It now turns out that $F$ is constant on $\scrH$ by the field equations~\cite{Ashtekar:1991vb}. By the above relation, so is then $F'$. An appropriate choice of $\alpha$ on $\scrH$ will then ensure any desired (constant) value for $F$ on $\scrH$. Following~\cite{Ashtekar:1991vb}, a completion with the choice
	\begin{equation}\label{F=1:H}
	 F~\hat{=}~1~,
	\end{equation}
will be called a~\emph{unit hyperboloid completion} (the name will be further justified below). We will exclusively work with such completions henceforth and often omit the qualifier `unit hyperboloid' when talking about them (or about the corresponding completion functions).

Away from $\scrH$, we may thus express the function $F$ as follows
	\begin{equation}\label{def:mu}
	 F = 1 + \mu\Omega~,
	\end{equation}
where $\mu$ is a function admitting a finite limit to $\scrH$. Following~\cite{Ashtekar:1991vb}, the function $\mu$ will be called the~\emph{mass aspect} of the spacetime $(\man, \met)$\footnote{The mass aspect has been denoted by ${}^{1}\!F$ in reference~\cite{Ashtekar:1991vb}.}. This quantity will play a very important role in our analysis, as we shall see later.

According to~\Eqref{F':H}, if the completion $(\hat{\man}, \Omega)$ is a unit hyperboloid completion, then the completion $(\hat{\man}, \Omega')$ will also be so provided
	\begin{equation}\label{alpha=1:H}
	 \alpha~\hat{=}~1~.
	\end{equation}
We will express the function $\alpha$ in $\nhood$ as
	\begin{equation}\label{def:alpha}
	 \alpha = 1 + \omega\Omega~,
	\end{equation}
where
	\begin{equation}\label{def:omega}
	 \omega~\equiv~\frac{\alpha - 1}{\Omega} = \frac{\Omega' - \Omega}{\Omega^{2}}~,
	\end{equation}
such that our analysis runs parallel with that of reference~\cite{Ashtekar:1991vb}\footnote{Note that the parametrization~\Eqref{def:alpha} of $\alpha$ in terms of $\omega$ does not cost us generality. This should be obvious when $\alpha$ and $\omega$ are smooth functions. However, even more generally, the parametrization~\Eqref{def:alpha} is simply another way to express definition~\Eqref{def:omega}. In particular, as discussed in the main text, every restriction on $\omega$ listed in~\Eqref{omega:sing} is in one-to-one correspondence with an analogous condition on $\alpha$.}. However, given our assumptions about $\alpha$, we cannot expect that $\omega$ is even continuous on $\scrH$. Rather, our goal is to determine how badly could $\omega$ be singular on $\scrH$ such that we can still meaningfully talk about unit hyperboloid completions.

We start by observing that since $\alpha$ is at least $C^1$ away from $\scrH$, $\omega$ must be so as well according to~\Eqref{def:omega}. Hence, in some appropriate neighborhood $\nhood$ of $\scrH$, we may express equations~\Eqref{dO':gen},~\Eqref{def:beta} and~\Eqref{n'^a:gen} in terms of $\omega$ and its derivatives as follows
	\begin{eqnarray}
	 \label{expr:dO'} & \Dl_{a}\Omega' = \beta\Dl_{a}\Omega + \Omega^{2}D_{a}\omega~, \\
	 \label{expr:beta} & \beta = 1 + 2\omega\Omega + \Omega^{2}\frac{\dl\omega}{\dl\Omega}~, \\
	 \label{expr:n'^a} & n^{\prime a} = \alpha^{-4}\left(\beta n^{a} + D^{a}\omega\right)~.
	\end{eqnarray}
Based on the analysis done so far we may thus infer:
	\begin{itemize}
	 \item Equations~\Eqref{alpha=1:H} and~\Eqref{def:alpha} imply $\omega\Omega~\hat{=}~0$.
	 \item The above fact, as well as $\Omega(\dl\alpha/\dl\Omega)~\hat{=}~0$ as noted earlier implies $\Omega^{2}(\dl\omega/\dl\Omega)~\hat{=}~0$\footnote{A more `covariantized' version of this requirement is that the function $\Omega^{2}\Lie_{n}\omega$ must vanish in the asymptotic limit. This follows from recalling definition~\Eqref{def:d/dO} and that the function $F$ admits a finite asymptotic limit~\Eqref{F=1:H}.\label{ftnt:Lien=ddO}}.
	 \item The one-form $D_{a}\omega$ must stay finite in the asymptotic limit in order for the vector $n^{\prime a}$ to stay finite on $\scrH$.
	\end{itemize}
For future reference, we will summarize these restrictions as follows: $\omega$ must be at least $C^{1}$ in $\nhood$ and satisfy
	 \begin{equation}\label{omega:sing}
	  \lim_{\Omega \to 0}\Omega\omega = 0~, \qquad\qquad \lim_{\Omega \to 0}\Omega^{2}\frac{\dl\omega}{\dl\Omega} = 0~, \qquad\qquad \lim_{\Omega \to 0}D_{a}\omega \neq \infty~.
	 \end{equation}
The above requirements thus imply the following asymptotic limits:
	\begin{eqnarray}
	 \label{dO':H} & \Dl_{a}\Omega'~\hat{=}~\Dl_{a}\Omega~, \\
	 \label{beta:H} & \beta~\hat{=}~1~, \\
	 \label{n'^a:H} & n^{\prime a}~\hat{=}~n^{a} + D^{a}\omega~.    
	\end{eqnarray}
\paragraph{Remarks:}
	\begin{itemize}
	 \item Equation~\Eqref{dO':H} shows that all unit hyperboloid completion functions must vanish on $\scrH$ at the `same rate' since the `normal' covectors $\Dl_{a}\Omega$ and $\Dl_{a}\Omega'$ agree on $\scrH$. This, in fact, is a `stronger' statement than these covectors being parallel on $\scrH$.
	 \item The above is not true for the corresponding `normal' vectors $n^{a}$ and $n^{\prime a}$, respectively according to equation~\Eqref{n'^a:H}; indeed, they are not even parallel. This is because, the physical metric being undefined on $\scrH$, the duality between the normal covectors and vectors are lost in the asymptotic limit. As we will see later, the difference between $n^{a}$ and $n^{\prime a}$ will provide a convenient way to describe the transformations between unit hyperbolic completions.
	 \item The expressions in~\Eqref{omega:sing} are all linear in $\omega$ (in the sense that if $\omega_{1}$ and $\omega_{2}$ are two functions satisfying these criteria, then so does the function $c_{1}\omega_{1} + c_{2}\omega_{2}$ for arbitrary constants $c_{1}$ and $c_{2}$).
	 \item While $\omega$ can indeed be singular in the asymptotic limit, the one-form $D_{a}\omega$ must~\emph{not} be so. This is somewhat counter-intuitive, since typically, the derivative of a function is usually more singular at a singularity of the function. This indicates that any `tangential derivative' of $\omega$ (\ie, a derivative along a direction parallel to the $\Omega =$ constant hypersurfaces) is `blind' to the singularity in the function.
	 \item If conditions~\Eqref{omega:sing} hold for some function (scalar) in some unit hyperboloid completion, then they also hold for the same function with respect to every unit hyperboloid completion. This `completion independence' property plays a key role in the context of gauge invariance discussed later\footnote{The notion of completion independence plays a very important role in our analysis. In~\ref{append:comp_indep}, we provide a general algorithm to establish completion independence of relevant quantities.}. 
	\end{itemize}
Our goal, now, is to find the most general version of $\omega$ satisfying these conditions. Our main result to that end is summarized in the following theorem:
	\begin{theorem}[Characterization of unit hyperboloid completion functions]\label{thm:omega}
Given a unit hyperboloid completion $(\hat{\man}, \Omega)$ of an asymptotically Minkowskian spacetime $(\man, \met)$, if a function $\omega$ is $C^{1}$ in a neighborhood $\nhood$ of $\scrH$ but (possibly) singular on $\scrH$ in a way such that the conditions in~\Eqref{omega:sing} hold, then the function $\Omega' \equiv (1 + \omega\Omega)\Omega$ defines another unit hyperboloid completion of $(\man, \met)$. Furthermore, any function $\omega$ satisfying the above conditions can always be expressed in the form
	 \begin{equation}\label{o=s+f}
	  \omega = \sigma + \omegaf~,
	 \end{equation}
where $\sigma$ is a function of $\Omega$ only which is at least $C^{1}$ in $\nhood$ but (possibly) singular on $\scrH$, satisfying
	 \begin{equation}\label{sigma:sing}
	  \lim_{\Omega \to 0}\Omega\sigma = 0~,
	 \end{equation}
as well as
	 \begin{equation}\label{sigma':sing}
	  \lim_{\Omega \to 0}\Omega^{2}\frac{\dd\sigma}{\dd\Omega} = 0~,
	 \end{equation}
and $\omegaf$ is also at least a $C^{1}$ function in $\nhood$ and satisfies
	 \begin{equation}\label{omegaf:sing}
	  \lim_{\Omega \to 0}\omegaf \neq \infty~, \qquad\qquad \lim_{\Omega \to 0}\Omega\frac{\dl\omegaf}{\dl\Omega} = 0~, \qquad\qquad \lim_{\Omega \to 0}D_{a}\omegaf \neq \infty~.
	 \end{equation}
	\end{theorem}
The condition~\Eqref{sigma:sing} implies condition~\Eqref{sigma':sing} via L'H\^opital rule applied to the ratio $\sigma/\Omega^{-1}$ in the asymptotic limit,~\emph{provided} that the function $\sigma$ satisfies certain additional criteria (\ie, that its derivative exists in $\nhood$ and that the quantity appearing on the left hand side of equation~\Eqref{sigma':sing} admits a finite limit asymptotically). By including~\Eqref{sigma':sing} in the statement of the above theorem, we are thus essentially assuming that the function $\sigma$ has all such required properties. Note that although the function $\omegaf$ is finite on $\scrH$ according to the first condition in~\Eqref{omegaf:sing}, its `$\Omega$-derivative' need not be so, as should be clear from the middle condition\footnote{We will soon realize that the middle condition in~\Eqref{omegaf:sing} does not guarantee a finite transformation of the mass aspect and that it needs to be strengthened further.}. A proof of Theorem~\ref{thm:omega} is presented in~\ref{proof:omega}.

We are now left with showing that the tensors $\qmet'_{a b}$ and $\qmet^{\prime a b}$ remain finite on $\scrH$ under change of completions. First, the relation between $\qmet'_{a b}$ and $\qmet_{a b}$ can be easily derived from the relevant definitions, and is given as follows
	\begin{equation}\label{expr:q'met_ab}
	 \qmet'_{a b} = \alpha^{2}\qmet_{a b} + \frac{\varpi\Dl_{a}\Omega\Dl_{b}\Omega}{\alpha^{2}F'F} - \frac{\beta(D_{a}\omega\Dl_{b}\Omega + D_{b}\omega\Dl_{a}\Omega)}{\alpha^{2}F'} - \frac{\Omega^{2}D_{a}\omega D_{b}\omega}{\alpha^{2}F'}~,
	\end{equation}
where
	\begin{equation*}
	 \varpi~\equiv~D_{c}\omega D^{c}\omega~.	    
	\end{equation*}
Both sides of equation~\Eqref{expr:q'met_ab} do admit finite limits to $\scrH$ as shown below
	\begin{equation}\label{q'met_ab:H}
	 \qmet'_{a b}~\hat{=}~\qmet_{a b} + \varpi\Dl_{a}\Omega\Dl_{b}\Omega - (D_{a}\omega\Dl_{b}\Omega + D_{b}\omega\Dl_{a}\Omega)~.
	\end{equation}
In a similar spirit, the analogous transformation rule between the tensors $\qmet^{\prime a b}$ and $\qmet^{a b}$ works out to be
	\begin{equation}\label{expr:q'met^ab}
	 \qmet^{\prime a b} = \frac{\qmet^{a b}}{\alpha^{2}} - \frac{\Omega^{2}}{\alpha^{6}F'}\left[\beta(n^{a}D^{b}\omega + n^{b}D^{a}\omega) + D^{a}\omega D^{b}\omega\right] + \frac{\Omega^{4}\varpi}{\alpha^{6}F'F}n^{a}n^{b}~.
	\end{equation}
Again, both sides of the above relation are expected to admit finite limits to $\scrH$, and they do. Taking the limits yield a much simpler relation than the corresponding covariant counterpart, namely
	\begin{equation}\label{q'met^ab:H}
	 \qmet^{\prime a b}~\hat{=}~\qmet^{a b}~.
	\end{equation}
\paragraph{Remarks:}
	\begin{itemize}
	 \item Both equations~\Eqref{q'met_ab:H} and~\Eqref{q'met^ab:H} are insensitive to the (possible) singularity in $\omega$.
	 \item Although $\qmet_{a b}~\hat{\neq}~\qmet'_{a b}$ they nevertheless induce the same three-metric on $\scrH$ (proof given in~\ref{proof:same_uhm_on_H}).
	 \item The asymptotic limit of the field equations dictate that the three metric induced is, in fact, the unit hyperboloid metric~\cite{Ashtekar:1991vb}. 
	\end{itemize}
These observations completely justify the terminology~\emph{unit hyperboloid completion}.

To sum up, we have shown how a change in a unit hyperboloid completion of the form~\Eqref{def:Omega'} transforms the kinematic quantities $\Dl_{a}\Omega$, $n^{a}$, $\qmet_{a b}$ and $\qmet^{a b}$ such that they all admit finite limits on $\scrH$ even in the presence of possible singularities in the function $\omega$ satisfying~\Eqref{omega:sing}. Apart from these kinematic objects there is another physical quantity, the mass aspect $\mu$ introduced in equation~\Eqref{def:mu}, whose asymptotic behavior is of importance because on $\scrH$ it encapsulates full information of the total $4$-momentum of the spacetime (see~\cite{Ashtekar:1991vb} and references therein). To derive it's transformation rule under~\Eqref{def:Omega'} we start from the relation between the functions $F$ and $F'$ appearing in equation~\Eqref{Fprime-alpha}. A straightforward calculation gives
	\begin{equation}\label{expr:mu}
	 \mu' = \alpha^{-5}\beta^{2}\mu + \alpha^{-5}(\beta + \alpha^{2})\varsigma + \cdots~,
	\end{equation}
where $\varsigma$ is given by
	\begin{equation}\label{def:varsigma}
	 \varsigma~\equiv~\Omega\left[\frac{\dd\sigma}{\dd\Omega} - \sigma^{2}\right],
	\end{equation}
and the `$\cdots$' represents terms which explicitly vanish in the asymptotic limit. Note, now, that $\varsigma$ can diverge on $\scrH$ even if $\sigma$ and its derivative satisfy the requirement of Theorem~\ref{thm:omega}. However, allowing this will completely undermine the physical significance of $\mu$. Rather, we must demand further restriction(s) on the function $\sigma$ (\ie, in addition to the requirements of Theorem~\ref{thm:omega}) such that $\mu$ may change at most by a finite constant under change in completions.

The additional restriction on $\sigma$ should be such that the following asymptotic limit holds
	\begin{equation}\label{varsigma:H}
	 \lim_{\Omega \to 0}\varsigma = \frac{\mu_{0}}{2}~,
	\end{equation}
where $\mu_{0}$ is a constant. This implies
	\begin{equation} \label{mu':H}
	 \mu'~\hat{=}~\mu + \mu_{0}~,
	\end{equation}
according to the transformation rule~\Eqref{expr:mu}. It turns out that the restrictions are easier to state in terms of the function $\Psi$ defined in terms of $\sigma$ as follows\footnote{The relation between $\sigma$ to $\Psi$ can be motivated as follows: The transformation rule~\Eqref{mu':H} suggests that the function $\varsigma'$ -- which is the analogue of the function $\varsigma$ under the transformation from $(\hat{\man}, \Omega')$ to $(\hat{\man}, \Omega)$ -- must approach the constant value $-\frac{1}{2}\mu_{0}$ asymptotically. It is also easy to see that asymptotically, the function $\omega'$ -- parametrizing the transformation from $(\hat{\man}, \Omega')$ to $(\hat{\man}, \Omega)$ -- becomes negative of the function $\omega$. Hence, the asymptotically significant contribution to $\varsigma$ must come from the $\Omega(\dd\sigma/\dd\Omega)$ term. This leads to the ansatz~\Eqref{def:Psi}.}
	\begin{equation}\label{def:Psi}
	 \Psi~\equiv~-\frac{\mu_{0}}{2} + \frac{\sigma}{\log\Omega}~.
	\end{equation}
Straightforward algebra then demonstrates that the condition~\Eqref{varsigma:H} will be ensured if $\Psi$ satisfies
	\begin{equation}\label{Psi:sing}
	 \lim_{\Omega \to 0}\Psi = 0~, \qquad\qquad \lim_{\Omega \to 0}\Omega\log\Omega\frac{\dd\Psi}{\dd\Omega} = 0~.
	\end{equation}
With these two conditions, we have ensured that the mass aspect remains well defined on $\scrH$ under any completion, but after having paid the price that the functional profile of the mass aspect has become completion sensitive.

Finally, to state our results in a completion independent fashion, we may introduce the function $\Theta$ via
	\begin{equation}\label{expr:omega}
	 \omega = \Theta\log\Omega~,
	\end{equation}
which is required to be at least $C^{1}$ in $\nhood$ and furthermore satisfy
	\begin{equation}\label{Theta:sing}
	 \lim_{\Omega \to 0}\Theta = \frac{\mu_{0}}{2}~, \qquad\qquad \lim_{\Omega \to 0}\Omega\log\Omega\frac{\dl\Theta}{\dl\Omega} = 0~, \qquad\qquad \lim_{\Omega \to 0}D_{a}\Theta = 0~.
	\end{equation}
It can easily be seen that the first two conditions on $\Theta$ implies 
	\begin{equation}\label{omega':sing}
	 \lim_{\Omega \to 0}\Omega\frac{\dl\omega}{\dl\Omega} = \frac{\mu_{0}}{2}~,
	\end{equation}
which is consistent with the middle condition in~\Eqref{omegaf:sing} but strictly stronger than the middle condition in~\Eqref{omega:sing}. Finally, both~\Eqref{Theta:sing} and~\Eqref{omega':sing} can be verified to be completion independent, a property that will play a key role in the context of gauge invariance (to be introduced later).

We wish to emphasize that the conditions~\Eqref{Theta:sing} are obtained by demanding finiteness of the kinematic quantities on $\scrH$. On the other hand, demanding finiteness of $\mu$ on $\scrH$ comes from its physical interpretation, which involves the dynamics of the underlying theory. This~\emph{physicality requirement} puts much more stringent conditions on the types of completions allowed.
\section{The group of supertranslations at spatial infinity}\label{sec:Str}
Let us now consider the set $\{\Omega\}$ of all unit hyperboloid completion functions. As we saw in the previous section, for every ordered pair $(\Omega', \Omega)$ there exits a function $\alpha$ which encodes the transformation between the completion functions via a rule of the form~\Eqref{def:Omega'}. We can, equivalently, describe the transformations in terms of the function $\omega$ which is related to $\alpha$ through equation~\Eqref{def:omega}.

In this section we will study the set of all transformations between unit hyperboloid completion functions. Let us denote this set by $\{\ST\}$. A sequence of two transformations from this set, say $\ST:\Omega \to \Omega' = \alpha\Omega$ and $\ST':\Omega' \to \Omega'' = \alpha'\Omega'$, will give us yet another transformation $\ST'':\Omega \to \Omega'' = \alpha''\Omega$, such that
	\begin{equation*}
	 \ST'' = \ST'\circ\ST \qquad\qquad\iim\qquad\qquad \alpha'' = \alpha'\alpha~,
	\end{equation*}
where `$\circ$' represents composition of maps in the usual sense. Thus, the set $\{\ST\}$ is closed under composition of the maps. Such maps clearly combine associatively due to associativity of usual multiplication between the `$\alpha$-functions'. It is then natural to ask whether a complete group structure exists on the set $\{\ST\}$.

To analyze the problem systematically, it will be necessary to have some way to identify the individual transformations. The most natural choice would be to label the transformations by the corresponding `$\omega$-function'. For example, the relation
	\begin{equation}\label{def:omega'}
	 \omega' = -\alpha^{-2}\omega~,
	\end{equation}
spelling out the `label' for the inverse transformation $\ST^{-1}:\Omega' \to \Omega$ justifies this expectation. Similarly, if one considers the sequence of transformations $\Omega \to \Omega' \to \Omega''$ discussed above, then the three associated `$\omega$-functions' can easily be seen to be related through
	\begin{equation}\label{def:omega''}
	 \omega'' = \omega + \omega' + 2\omega\omega'\Omega + \omega^{2}\omega'\Omega^{2}~. 
	\end{equation}
Note that~\Eqref{def:omega''} would have been symmetric under the exchange of $\omega$ and $\omega'$ without the very last term on the right hand side. The physical significance of this `asymmetry inducing' term will be discussed shortly.

On the other hand, there will exist transformations which relate completion functions that may be distinct in the bulk of the spacetime but are `essentially indistinguishable close to $\scrH$'. These are the~\emph{gauge transformations} in the present context. Clearly, the precise notion of `asymptotic indistinguishability' or~\emph{gauge equivalence} between two completion functions should depend on the asymptotic properties of the associated `$\omega$-function'. Following reference~\cite{Ashtekar:1991vb},~\emph{we will regard two completion functions to be gauge equivalent if the `$\omega$-function' associated with their ordered pair vanishes asymptotically}.

We can easily demonstrate that this notion of gauge equivalence, to be denoted by `$\cong$' henceforth, defines a genuine equivalence relation on the set $\{\Omega\}$. As with any set with an equivalence relation defined on, the relation `$\cong$' partitions the set $\{\Omega\}$ into~\emph{disjoint} equivalence classes. We will denote the equivalence class containing the function $\Omega$ by $\ripple{\Omega}$, and refer to it as a~\emph{ripple} on the asymptotic geometry~\emph{\`a la} reference~\cite{Ashtekar:1991vb}. Thus, gauge equivalence between two completion functions $\Omega$ and $\Omega'$ can be denoted by $\Omega' \cong \Omega$, and two such functions will be members of the ripple $\ripple{\Omega}$ (or equivalently, of the ripple $\ripple{\Omega'}$).
	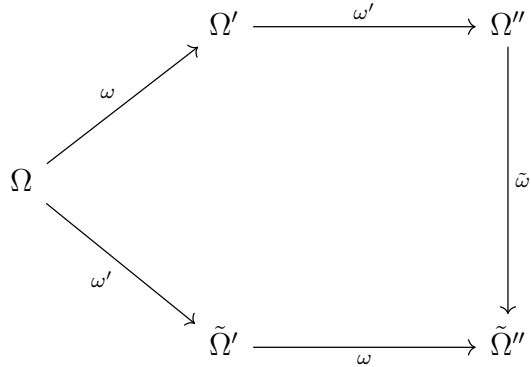
\begin{figure}
	 \begin{center}
	  \begin{tikzcd}
	   && {{\Omega'}} &&& {{\Omega''}} \\
	   \\
	   {{\Omega}} \\
	   \\
	   && {{\tilde{\Omega}'}} &&& {{\tilde{\Omega}''}}
	   \arrow["{{\omega'}}", from=1-3, to=1-6]
	   \arrow["{{\tilde{\omega}}}", from=1-6, to=5-6]
	   \arrow["{{\omega}}", from=3-1, to=1-3]
	   \arrow["{{\omega'}}"', from=3-1, to=5-3]
	   \arrow["{{\omega}}"', from=5-3, to=5-6]
	  \end{tikzcd}
	  \caption{Schematic diagram of transformations between completion functions. Each directed arrow represents a transformation and the label shows the corresponding `$\omega$-function'.}
	  \label{commdiag}
	 \end{center}
	\end{figure}
Consider now the following two sequences of transformations given in Figure~\Eqref{commdiag}. If the functions $\Omega''$ and $\tilde{\Omega}''$ are gauge equivalent, then the~\emph{order} of the two transformations acting on $\Omega$ is also physically irrelevant asymptotically (\ie, the two transformations will~\emph{commute} asymptotically). This is indeed what happens if the `$\omega$-functions' are smooth~\cite{Ashtekar:1991vb}, but the situation is certainly not clear yet in the present setting.

Using the formula~\Eqref{def:omega''}, the function $\tilde{\omega}$ in Figure~\Eqref{commdiag} can easily be determined to be
	\begin{equation*}
	 \tilde{\omega}~\propto~\omega\omega'(\omega' - \omega)\Omega^{2}~,
	\end{equation*}
where the proportionality factor goes to unity asymptotically. Note that the expression on the right hand side is a consequence of the `asymmetry inducing' terms present in the formula~\Eqref{def:omega''}. Gauge (in)equivalence of $\Omega''$ and $\tilde{\Omega}''$ is determined by checking whether the function $\tilde{\omega}$ vanishes asymptotically.

Remarkably, while $\tilde{\omega}$ need not necessarily vanish in the asymptotic limit if $\omega$ and $\omega'$ are merely restricted by the first condition in~\Eqref{omega:sing}, it~\emph{does} vanish asymptotically if both $\omega$ and $\omega'$ are expressed as in equation~\Eqref{expr:omega} with the corresponding `$\Theta$-functions' which satisfy~\Eqref{Theta:sing}. Consequently, all transformation in the set $\{\ST\}$ are mutually commutative asymptotically, just as in the special case of $\omega$ being smooth, provided the mass aspects associated with all the completions under consideration are finite. This indicates:
	\begin{itemize}
	 \item the dynamical information encoded in $\mu$ plays a significant role in gauge (in)equivalence of the completion functions;
	 \item it is essential to phrase gauge equivalence in terms of the function $\Theta$.
	\end{itemize}

To that end, let $\bTheta$ be the set of all (real) functions satisfying the conditions in~\Eqref{Theta:sing}, and let $\bThetao$ be a proper subset of $\bTheta$ consisting of those functions for which the following condition holds additionally 
	\begin{equation}\label{Theta:gauge}
	 \lim_{\Omega \to 0}\Theta\log\Omega = 0~.
	\end{equation}
Based on the identification~\Eqref{expr:omega}, The above condition is clearly a restatement of asymptotic vanishing of some `$\omega$-function'. In other words, any function $\Theta \in \bThetao$ describes a transformation between two gauge equivalent completion functions. Since the conditions~\Eqref{Theta:sing} and~\Eqref{Theta:gauge} are linear in $\Theta$, both $\bTheta$ and $\bThetao$ are closed under linear combinations of their respective elements. Equally importantly, the gauge condition~\Eqref{Theta:gauge} is manifestly independent of the choice of the unit hyperboloid completion (in spite of the explicit appearance of the $\log\Omega$ factor).

Two functions $\Theta, \Theta' \in \bTheta$ will be regarded as~\emph{asymptotically equivalent} if $\Theta' - \Theta \in \bThetao$\footnote{Asymptotic equivalence, like gauge equivalence, is an equivalence relation. While they are very closely related (two asymptotically equivalent members of $\bTheta$ generates the same~\emph{physical} transformation), it is nevertheless important to acknowledge their difference. The latter concept is relevant for completion functions, while the former concept pertains to the elements of the set $\bTheta$ describing transformations between the completion functions.}. We will employ the following notation to denote asymptotic equivalence between members of $\bTheta$
	\begin{equation}\label{Theta:asymp_equiv}
	 \Theta' = \eqv{\Theta} \qquad \iim\quad \Theta = \eqv{\Theta'} \qquad\iim\quad\qquad \Theta' - \Theta \in \bThetao~.
	\end{equation}
By the above definition, every function $\Theta$ is asymptotically equivalent to itself. In the same spirit we may represent every element of the set $\bThetao$ as $\eqv{0}$. Also, if the function $\Theta$ induces a transformation between two completion functions then the inverse transformation, according to~\Eqref{def:omega'}, will be induced by the function $-\eqv{\Theta}$.

These new notational tools now help us to succinctly summarize properties of the transformations in a manifestly gauge invariant way as follows:
	\begin{itemize}
	 \item We label elements of $\{\ST\}$ as $\ST_{\eqv{\Theta}}$.
	 \item The statement that gauge transformations keep completion functions within the same ripple becomes
	  \begin{equation*}
	   \ST_{\eqv{0}}\ripple{\Omega} = \ripple{\Omega}~.
	  \end{equation*}
	 \item The effect of physical transformations are represented by
	  \begin{equation*}
	   \ST_{\eqv{\Theta}}\ripple{\Omega} = \ripple{\Omega'}~.
	  \end{equation*}
	 \item The inverse of a transformation, based on previous observations, is
	  \begin{equation*}
	   \ST_{-\eqv{\Theta}}\ripple{\Omega'} = \ripple{\Omega}~.
	  \end{equation*}
	 \item The transformations commute, \ie,
	  \begin{equation*}
	   \ST_{\eqv{\Theta'}}\circ\ST_{\eqv{\Theta}} = \ST_{\eqv{\Theta}}\circ\ST_{\eqv{\Theta'}} = \ST_{\eqv{\Theta + \Theta'}}~.
	  \end{equation*}
	\end{itemize}
Therefore, if we identify $\ST_{\eqv{0}}$ as the identity element in the set, along with the element $\ST_{-\eqv{\Theta}}$ as the inverse of $\ST_{\eqv{\Theta}}$ for each $\Theta$, then the set\footnote{Reference~\cite{Ashtekar:1991vb} denotes this group by $\CMcal{S}$.}
	\begin{equation}\label{def:Str}
	 \Str~\equiv~\left\{\ST_{\eqv{\Theta}}~|~\Theta\,\in\,\bTheta\right\}~,
	\end{equation}
can indeed be regarded as an infinite dimensional Abelian group,~\emph{isomorphic} to the group describing the linear combination of functions in $\bTheta$ modulo elements in $\bThetao$. For reasons that will become clear eventually, this group will henceforth be called~\emph{the group of supertranslations at spatial infinity}. Since the set $\bTheta$ is insensitive to the unit hyperboloid completions as noted above, the group $\Str$ is so as well and therefore describes physical transformations in the spacetime.

While the notion of the group of supertranslations introduced above seems to be a minor generalization of the analogous concept appearing in~\cite{Ashtekar:1991vb}, there is in fact a significant conceptual difference between the descriptions of the two constructs. For example, two functions $\Theta, \Theta' \in \bTheta$ can approach the same constant value on $\scrH$ (by the first condition in~\Eqref{Theta:sing}) and yet their difference (which must vanish by the above assumption) may not satisfy the gauge condition~\Eqref{Theta:gauge}. Hence, two such functions will represent two physically distinct transformations $\ST_{\eqv{\Theta}}$ and $\ST_{\eqv{\Theta'}}$, respectively. This example highlights the fact that the behavior of the functions in $\bTheta$ in the bulk of the spacetime is physically significant. This is not true in the way the group of supertranslations is described in~\cite{Ashtekar:1991vb}, where only the profile of the (smooth) function $\omega$ on $\scrH$ is relevant.

Recall that the boundary $\scrH$ and the unit hyperboloid metric induced on $\scrH$ stay invariant by Definition~\ref{def:AR:asymp}, and so do the one-form $\Dl_{a}\Omega$ as well as the $(2, 0)$-tensor $\qmet^{a b}$ according to the transformation rules~\Eqref{dO':H} and~\Eqref{q'met^ab:H}, respectively (stated differently, the one-form $\Dl_{a}\Omega$ and the tensor $\qmet^{a b}$ associated with gauge inequivalent ripples agree on $\scrH$). In this sense that these structures can be regarded as~\emph{rigid}. On the other hand, the vector $n^{a}$ and the $(0, 2)$-tensor $\qmet_{a b}$ transform according to the rules~\Eqref{n'^a:H} and~\Eqref{q'met_ab:H}, respectively. Consequently, two physically distinct ripples will have different $n^{a}$-vectors and $\qmet_{a b}$-tensors associated with them (although the latter will induce the same unit hyperboloid metric on $\scrH$).

Of course,  all of the above-mentioned kinematic structures remain invariant under gauge transformations. Indeed this is~\emph{the} feature that distinguishes gauge transformations from physical transformations. When we talk about asymptotic symmetries and their generators in the next section, this observation will play a key role in identifying the corresponding generators. With this as the motivation, we propose the following definition:
\begin{definition}[Universal structure of an AM spacetime]\label{def:USAM}
The universal structure of an asymptotically Minkowskian spacetime, in accordance with Definition~\ref{def:AR:AM}, is the collection of the following substructures:
	\begin{enumerate}
	 \item a~\underbar{rigid} substructure consisting of 
	  \begin{enumerate}
	   \item the (topologically $\RR \times \Sph^{2}$ and geodesically complete) boundary $\scrH$;
	   \item the unit hyperboloid metric induced on $\scrH$;
	   \item the one-form $\Dl_{a}\Omega$ on $\scrH$;
	   \item the rank-$(2, 0)$ symmetric tensor $\qmet^{a b}$ on $\scrH$;
	  \end{enumerate}
	 \item the set of all the~\underbar{ripples} $\{\ripple{\Omega}\}$ on the asymptotic geometry.
	\end{enumerate}
\end{definition}
Note that the possibility of the function $\omega$ to be asymptotically divergent makes the universal structure here `weaker' than that considered in~\cite{Ashtekar:1991vb}.
\section{Asymptotic symmetries of AM spacetimes}\label{sec:ASAM}
According to Definition~\ref{def:USAM}, the universal structure associated with any asymptotically Minkowskian spacetime consists of a rigid structure and a collection of ripples. We may formally define an~\emph{asymptotic symmetry transformation} as a diffeomorphism (acting on the fictitious manifold $\hat{\man}$) which leaves the rigid substructure invariant but may affect the ripples. The exhaustive list of possibilities are:
	\begin{itemize}
	 \item The rigid structure remains trivially unaffected but the ripples transform to inequivalent ripples. Example: transformations between unit hyperboloid completions discussed in the previous section.
	 \item Both the rigid substructure and the ripples are unaffected. Example: gauge transformations, also discussed previously.
	 \item Transformation reduces to an~\emph{isometry} of the unit hyperboloid metric induced on $\scrH$. In this case, the rigid substructure also stays invariant, but not trivially.
	\end{itemize}
We expect that the most general diffeomorphism leaving the universal structure invariant should be a composition of successive `infinitesimal' transformations which individually have one of these effects. We will demonstrate below that this intuition is indeed correct.

Henceforth, we will call the Lie group of all asymptotic symmetry transformations as the~\emph{Spi group}\footnote{\emph{\`A la} reference~\cite{Ashtekar:1991vb}, although they use the notation $\CMcal{G}$ for the Spi group.}. Of course, this group will be `larger' than the group $\Str$ since it will include transformations which keep the rigid structure invariant non-trivially. Naturally, the first step is to study the Lie algebra $\spi$ associated with $\Spi$. The Lie algebra $\spi$ is generated by vector fields called the~\emph{asymptotic symmetry generators}. Informally, they can be thought of as vector fields generating `infinitesimal versions' of the asymptotic symmetries.

A vector field generating an asymptotic symmetry is expected to admit a finite asymptotic limit, since the diffeomorphism it generates must affect the asymptotic boundary along with the fields on it. However, our analysis of transformations between unit hyperboloid completions suggest that we should~\emph{not} expect such fields to necessarily admit~\emph{smooth} limits to $\scrH$. We will address this issue by following the guideline below:
	\begin{itemize}
	 \item retain as much of the properties of the smooth generators (\eg, those studied in reference~\cite{Ashtekar:1991vb}) as possible, without relying too much on their smoothness;
	 \item appeal to the general intuition that the Lie derivative of any tensor along a given vector field describes, heuristically, an `infinitesimal change' of the tensor under the diffeomorphism generated by the vector field;
	 \item leverage the observation that a prerequisite to preservation of the universal structure is the preservation of the boundary $\scrH$. 
	\end{itemize}
With these insights, we propose the following definition:
\begin{definition}[$\scrH$-preserving vector fields of AM spacetimes]\label{def:HPVF}
A vector field $X^{a}$ in an asymptotically Minkowskian spacetime $(\man, \met)$ will be regarded as $\scrH$-preserving with respect to a unit hyperboloid completion $(\hat{\man}, \Omega)$ if $X^{a}$ admits a finite limit to $\scrH$, and furthermore, the asymptotic limits of the tensor fields $\Omega^{-1}\Lie_{X}\Omega$, $\Lie_{X}\Dl_{a}\Omega$, $\Lie_{X}n^{a}$ and $\Lie_{X}\qmet^{a b}$ are all finite.
\end{definition}
In other words, a vector field is $\scrH$-preserving if (i) it exists asymptotically; (ii) one can meaningfully talk about the asymptotic changes (induced by the diffeomorphism it generates) to the available geometric structure on $\scrH$; and (iii) the boundary $\scrH$ is preserved under this diffeomorphism (in a sense to be explained shortly). We can verify that $\Lie_{X}\qmet_{a b}$ admits a finite asymptotic limit if the conditions of the above definition hold.

It turns out, however, that a vector field satisfying the requirements of the above definition with respect to a given completion need not do the same with respect to another (inequivalent) completion. Rather, to achieve that goal, we also need the function $\Theta$ to be at least $C^{2}$ in $\nhood$ as well as satisfy
	\begin{equation}\label{Theta'':sing}
	 \lim_{\Omega \to 0}\Omega^{2}\log\Omega\frac{\dl^{2}\Theta}{\dl\Omega^{2}} = 0~,
	\end{equation}
in addition to the conditions listed in equation~\Eqref{Theta:sing}, and also require the asymptotic limit of the function $\omegaf$ to be at least $C^{2}$~\emph{on} $\scrH$\footnote{One may show that the condition~\Eqref{Theta'':sing} implies that the function $\omegaf$ must also be at least $C^{2}$ in $\nhood$ and satisfies
	\begin{equation*}
	 \lim_{\Omega \to 0}\Omega^{2}\left(\frac{\dl^{2}\omegaf}{\dl\Omega^{2}}\right) = 0~.
	\end{equation*}
The above should hold along with the conditions listed in equations~\Eqref{omegaf:sing}.}. Under such restrictions, we then have the desired result in the form of the following theorem:
	\begin{theorem}[Completion independence of $\scrH$-preserving vector fields]\label{thm:HPVF}
If $X^{a}$ is a $\scrH$-preserving vector field according to Definition~\ref{def:HPVF} with respect to a given unit hyperboloid completion, then it is also a $\scrH$-preserving vector field with respect to every unit hyperboloid completion, provided that the group of supertranslations relating such completions is restricted by conditions~\Eqref{Theta:sing} and~\Eqref{Theta'':sing} as well as that the asymptotic limit of the function $\omegaf$ be at least $C^{2}$ on $\scrH$.
	\end{theorem}
A proof of the above theorem amounts to verifying the existence of the necessary limits; the essential details appear in~\ref{proof:HPVF}.

According to the above theorem, a $\scrH$-preserving vector field $X^{a}$ will satisfy
	\begin{equation}\label{def:u}
	 \Lie_{X}\Omega = u\Omega~,
	\end{equation}
with respect to any completion function $\Omega$, where the function $u \equiv \Omega^{-1}X^{a}\Dl_{a}\Omega$ admits a finite asymptotic limit. Consequently, $\Lie_{X}\Omega$ itself will vanish asymptotically, and this must be true for every completion function. This means that irrespective of the completion, the `$\Omega = 0$' surface is always `held back in its original place' under the diffeomorphism generated by $X^{a}$. In other words,~\emph{the diffeomorphism generated by a $\scrH$-preserving vector field keeps the boundary $\scrH$ invariant}. This justifies their name.

We can go even further and define an orthogonal decomposition of $X^{a}$ with respect to the completion $(\hat{\man}, \Omega)$ using
	\begin{equation}\label{orthodecomparb}
	 U^{a}~\equiv~\frac{\Omega u}{F}n^{a}~, \qquad\qquad V^{a}~\equiv~X^{a} - U^{a}~,
	\end{equation}
These vectors are  are also $\scrH$-preserving according to Theorem~\ref{thm:HPVF}. Finally, if $X_{1}^{a}$ and $X_{2}^{a}$ are two $\scrH$-preserving vector fields, then linearity of Lie derivatives will ensure that any arbitrary linear combination $X^{a} = c_{1}X_{1}^{a} + c_{2}X_{2}^{a}$, for arbitrary constant $c_{1}$ and $c_{2}$, will also be one.

We may now define an asymptotic symmetry generator of an asymptotically Minkowskian spacetime as follows:
\begin{definition}[Asymptotic symmetry generator of an AM spacetime]\label{def:ASAM}
A $\scrH$-preserving vector field $\xi^{a}$, as per Theorem~\ref{thm:HPVF}, will be called an asymptotic symmetry generator with respect to a unit hyperboloid completion $(\hat{\man}, \Omega)$ if it keeps the one-form $\Dl_{a}\Omega$, the rank-$(2, 0)$ tensor $\qmet^{a b}$ and the inverse physical metric~\emph{asymptotically invariant} in the following sense:
	\begin{eqnarray}
	 \label{def:ASAM:Lxi_dO:H} \lim_{\Omega \to 0}\Lie_{\xi}\Dl_{a}\Omega = 0~, \\
	 \label{def:ASAM:Lxi_qmet^ab:H} \lim_{\Omega \to 0}\Lie_{\xi}\qmet^{a b} = 0~, \\
	 \label{def:ASAM:AIIPM} \lim_{\Omega \to 0}\Omega^{-4}(\Lie_{\xi}\met^{a b})\Dl_{b}\Omega = 0~.
	\end{eqnarray}
\end{definition}
\paragraph{Remarks:}
	\begin{itemize}
	 \item Conditions~\Eqref{def:ASAM:Lxi_dO:H} and~\Eqref{def:ASAM:Lxi_qmet^ab:H} are necessary to preserve the rigid substructure introduced in Definition~\ref{def:USAM}. Although they are similar to the transformation rules~\Eqref{dO':H} and~\Eqref{q'met^ab:H}, respectively, these conditions are much more general since these apply to~\emph{all} kinds of asymptotic symmetry generators, not just those generating changes of the unit hyperboloid completion.
	 \item As will be shown below, the condition~\Eqref{def:ASAM:AIIPM} captures the action of an asymptotic symmetry generator on the ripples on the asymptotic geometry (this can be anticipated by noting that the condition implies an equivalent condition on the vector $\Lie_{\xi}n^{a}$ via equation~\Eqref{def:n^a}). Heuristically, this condition also forces an asymptotic symmetry generator to behave like a Killing vector asymptotically.
	\end{itemize}
Our job, now, is to construct the most general vector $\xi^{a}$ satisfying the requirements of Definition~\ref{def:ASAM}. To that end, our first major result is stated as the following theorem:
	\begin{theorem}[Characterization of asymptotic symmetry generators]\label{thm:w}
Let $\xi^{a}$ be an asymptotic symmetry generator with respect to a unit hyperboloid completion $(\hat{\man}, \Omega)$ as per Definition~\ref{def:ASAM}. Then the function
	 \begin{equation}\label{def:w}
	  w~\equiv~\Omega^{-2}\Lie_{\xi}\Omega~,
	 \end{equation}
must be at least $C^{1}$ in $\nhood$, and while it may be singular asymptotically, its divergence on $\scrH$ will be restricted by the following conditions
	 \begin{equation}\label{w:sing}
	  \lim_{\Omega \to 0}\Omega w = 0~, \qquad\qquad \lim_{\Omega \to 0}\Omega^{2}\frac{\dl w}{\dl\Omega} = 0~, \qquad\qquad \lim_{\Omega \to 0}D_{a}w \neq \infty~,
	 \end{equation}
as a consequence of requirements~\Eqref{def:ASAM:Lxi_dO:H} and~\Eqref{def:ASAM:AIIPM}.
	\end{theorem}
A simple proof of the above theorem appears in~\ref{proof:w}. Remarkably, conditions~\Eqref{w:sing} are completely analogous to those in~\Eqref{omega:sing}. This strongly suggests that an asymptotic symmetry generator (with a non-trivial profile for the function $w$ in a neighborhood $\nhood$ of $\scrH$) induces some analogue of the transformations between unit hyperboloid completions (but possibly implementing other transformations along with it). Our next goal will be to explore the properties of general asymptotic symmetry generators in sufficient details in order to address these as well as other issues concerning them.
\subsection{Asymptotic symmetry generators: basic properties}\label{sec:ASAM:props}
An asymptotic symmetry generator $\xi^{a}$ can always be  orthogonally decomposed with respect to the $\Omega =$ constant hypersurfaces as follows
	\begin{equation}\label{def:taun^a}
	 \xi^{a}~\equiv~\taun^{a} + \Xi^{a}~, \qquad\qquad\qquad \taun^{a}~\equiv~\frac{\Omega^{2}w}{F}n^{a}~,
	\end{equation}
where the function $w$ is defined as in equation~\Eqref{def:w}, and the vector field $\Xi^{a}$ is $\scrH$-preserving and satisfies $\Xi^{a}\Dl_{a}\Omega = 0$. Furthermore, as a consequence of requirement~\Eqref{def:ASAM:AIIPM}, $\xi^{a}$ satisfies
	\begin{equation}\label{Lxi_n^a:H}
	 \lim_{\Omega \to 0}\Omega^{-4}(\Lie_{\xi}\met^{a b})\Dl_{b}\Omega = 0 \qquad\qquad\iim\qquad\qquad \Lie_{\xi}n^{a}~\hat{=}~D^{a}w~.
	\end{equation}
The above relation thus captures the effect of the generator $\xi^{a}$ on the ripples on the asymptotic geometry exactly as anticipated before.

From the definition of the vector $\taun^{a}$ introduced above, we may further deduce
	\begin{eqnarray}
	 \label{Ltaun_dO:H} \Lie_{\taun}\Dl_{a}\Omega~\hat{=}~0~, \\ 
	 \Lie_{\taun}n^{a}~\hat{=}~0~, \\
	 \lim_{\Omega \to 0}\Omega^{-4}(\Lie_{\taun}\met^{a b})\Dl_{b}\Omega~=-D^{a}w~, \\
	 \label{Ltaun_qmet^ab:H} \Lie_{\taun}\qmet^{a b}~\hat{=}~0~. 	    
	\end{eqnarray}
Thus, although $\taun^{a}$ is $\scrH$-preserving by construction, it cannot qualify as an asymptotic symmetry generator since it does not preserve the inverse physical metric asymptotically (unless, of course, $w$ is a constant).

For reasons to be clear very soon, let us also define the vector $\taup^{a}$ as follows
	\begin{equation}\label{def:taup^a}
	 \taup^{a}~\equiv~-\Omega D^{a}w~.
	\end{equation}
Thus $\taup^{a}$ is $\scrH$-preserving by construction and it vanishes on $\scrH$. Simple calculations furthermore demonstrate
	\begin{eqnarray}
	 \label{Ltaup_dO:H} \Lie_{\taup}\Dl_{a}\Omega~\hat{=}~0~, \\
	 \Lie_{\taup}n^{a}~\hat{=}~D^{a}w~,\\
	 \lim_{\Omega \to 0}\Omega^{-4}(\Lie_{\taup}\met^{a b})\Dl_{b}\Omega =~D^{a}w~, \\ 
	 \label{Ltaup_qmet^ab:H} \Lie_{\taup}\qmet^{a b}~\hat{=}~0~. 
	\end{eqnarray}
Clearly, $\taup^{a}$ cannot be an asymptotic symmetry generator either (in fact, $\taup^{a}$ vanishes in the bulk of the spacetime if $w$ becomes a constant in $\scrH$ and it vanishes on $\scrH$ under all circumstances).

However, consider the vector $\tau^{a}$ defined as follows
	\begin{eqnarray}\label{def:tau^a}
	 \tau^{a} &~\equiv~& \taun^{a} + \taup^{a}~, \nonumber \\
                  &~=~& \frac{\Omega^{2}w}{F}n^{a} - \Omega D^{a}w~.
	\end{eqnarray}
By definition, $\tau^{a}$ is $\scrH$-preserving, it vanishes on $\scrH$ and satisfies
	\begin{equation}\label{expr:Ltau_O}
	 \Lie_{\tau}\Omega = \Omega^{2}w~.
	\end{equation}
Using linearity of limits and Lie derivatives, we can read off from~\Eqref{Ltaun_dO:H} --~\Eqref{Ltaun_qmet^ab:H} and~\Eqref{Ltaup_dO:H} --~\Eqref{Ltaup_qmet^ab:H}
	\begin{eqnarray}
	 \label{Ltau_dO:H} \Lie_{\tau}\Dl_{a}\Omega~\hat{=}~0~, \\
	 \Lie_{\tau}n^{a}~\hat{=}~D^{a}w~, \\
	 \lim_{\Omega \to 0}\Omega^{-4}(\Lie_{\tau}\met^{a b})\Dl_{b}\Omega =~0~, \\
	 \label{Ltau_qmet^ab:H}  \Lie_{\tau}\qmet^{a b}~\hat{=}~0~.
	\end{eqnarray}
Thus $\tau^{a}$, unlike its constituents,~\emph{does} satisfy all the defining properties of an asymptotic symmetry generator. Hence it is an asymptotic symmetry generator itself. Note that in a `Beig-Schmit'-type coordinate system with $\Omega$ as one of the coordinate functions (see, \eg, references~\cite{Beig:1982ifu, Ashtekar:1991vb, Compere:2011ve, Troessaert:2017jcm}), the `$\Omega$-component' of $\tau^{a}$ is just $\Omega^{2}w$ according to the identification~\Eqref{def:d/dO}.

Now, let the vector $\psi^{a}$ denote the `left-over bit' in $\Xi^{a}$ upon the `removal' of $\taup^{a}$ from it, \ie,
	\begin{equation}\label{def:psi^a}
	 \psi^{a}~\equiv~\Xi^{a} - \taup^{a}~,
	\end{equation}
such that from equations~\Eqref{def:taun^a} and~\Eqref{def:tau^a}
	\begin{equation*}
	 \xi^{a} = \tau^{a} + \psi^{a}~.
	\end{equation*}
We will now demonstrate that $\psi^{a}$ is also an asymptotic symmetry generator.

To that end, note that by construction, $\psi^{a}$ is $\scrH$-preserving and orthogonal to $\Dl_{a}\Omega$, \ie,
	\begin{equation}\label{expr:Lpsi_O}
	 \psi^{a}\Dl_{a}\Omega = 0 \qquad\qquad\iim\qquad\qquad \Lie_{\psi}\Omega = 0~.
	\end{equation}
As a consequence of the above we then also have
	\begin{equation}\label{expr:Lpsi_dO}
	 \Lie_{\psi}\Dl_{a}\Omega = 0~.
	\end{equation}
Invoking linearity of limits and Lie derivatives yet again appropriately, we may furthermore infer
	\begin{equation}\label{Lpsi_n^a:H}
	 \lim_{\Omega \to 0}\Omega^{-4}(\Lie_{\psi}\met^{a b})\Dl_{b}\Omega = 0 \qquad\qquad\iim\qquad\qquad \Lie_{\psi}n^{a}~\hat{=}~0~.
	\end{equation}
Finally, since $\xi^{a}$ and $\tau^{a}$ are both asymptotic symmetry generators (former by assumption, latter as has just been explicitly demonstrated above), linearity of limits and Lie derivatives dictate
	\begin{equation}\label{Lpsi_qmet^ab:H}
	 \Lie_{\psi}\qmet^{a b}~\hat{=}~0~.
	\end{equation}
Therefore, $\psi^{a}$ must also be an asymptotic symmetry generator, as promised. However, since $\psi^{a}$ does not vanish asymptotically, it must then satisfy the Killing equation on $\scrH$, \ie
	\begin{equation}\label{KVE:psi:H}
	 D_{a}\psi_{b} + D_{b}\psi_{a}~\hat{=}~0~.
	\end{equation}
Consequently, the diffeomorphism generated by $\psi^{a}$ also preserves the universal structure (albeit non-trivially).

The general results used in the proof of Theorem~\ref{thm:HPVF} applied to the above also yield
	\begin{eqnarray} \label{Lxi_qmet_ab:H}
	  \Lie_{\tau}\qmet_{a b} &~\hat{=}~-(D_{a}w)\Dl_{b}\Omega - (D_{b}w)\Dl_{a}\Omega~, \\
	  \Lie_{\psi}\qmet_{a b} &~\hat{=}~0~.
	\end{eqnarray}
By linearity of limits and Lie derivatives, the analogous result for $\Lie_{\xi}\qmet_{a b}$ is therefore identical with that for $\Lie_{\tau}\qmet_{a b}$.

This concludes our presentation of the asymptotic limits of the actions of the asymptotic symmetry generators on all the `kinematic' fields associated with the completion $(\hat{\man}, \Omega)$. In particular, the results in equations~\Eqref{Lxi_n^a:H} and~\Eqref{Lxi_qmet_ab:H} confirm that the definition of asymptotic symmetry generators proposed in the current work is indeed an appropriate generalization of that appearing in reference~\cite{Ashtekar:1991vb} (see equation~(4.3) of that reference in particular)\footnote{We should also note that while the decomposition~\Eqref{xi=tau+psi} (obviously) holds for the smooth generators studied in reference~\cite{Ashtekar:1991vb}, the convention followed in that reference to denote the asymptotic limit of the decomposition by an ordered pair -- consisting of the asymptotic profile of the function $w$ and the asymptotic limit of the vector $\psi^{a}$ -- is not appropriate any more, since the former quantity is divergent in general.}. The main take-away message from the above analysis is summarized in the form of the following theorem:
	\begin{theorem}[Basic characterization of asymptotic symmetry generators]\label{thm:ASAM:props}
An asymptotic symmetry generator $\xi^{a}$ with respect to a unit hyperboloid completion $(\hat{\man}, \Omega)$, as per Definition~\ref{def:ASAM}, can always be represented as
	\begin{equation}\label{xi=tau+psi}
	 \xi^{a} = \tau^{a} + \psi^{a}~,
	\end{equation}
where the asymptotic symmetry generator $\tau^{a}$~\Eqref{def:tau^a} is constructed out of the function $w$~\Eqref{def:w} which satisfies the restrictions of Theorem~\ref{thm:w}, and the asymptotic symmetry generator $\psi^{a}$ (which is tangential to the $\Omega =$\,constant surfaces everywhere) becomes a Killing vector of the unit hyperboloid metric on $\scrH$ asymptotically~\Eqref{KVE:psi:H}. Furthermore, $\xi^{a}$ approaches $\psi^{a}$ asymptotically since $\tau^{a}$ vanishes in the same limit by construction. Consequently, $\xi^{a}$ also acts as a Killing vector of the unit hyperboloid metric on $\scrH$.
	\end{theorem}
As a complement to the above theorem, the next theorem lays down the necessary conditions that any given vector field needs to satisfy in order to qualify as an asymptotic symmetry generator:
	\begin{theorem}[Necessary properties for asymptotic symmetry generators]\label{thm:ASAM:identify}
A vector field $\xi^{a}$ in an asymptotically Minkowski spacetime $(\man, \met)$ will be an asymptotic symmetry generator with respect to a unit hyperboloid completion $(\hat{\man}, \Omega)$ if
	 \begin{enumerate}
	  \item $\xi^{a}$ is a $\scrH$-preserving vector field;
	  \item the scalar $w \equiv \Omega^{-2}\Lie_{\xi}\Omega$ satisfies the conditions of Theorem~\ref{thm:w};
	  \item the vector field $\psi^{a} \equiv \xi^{a} - F^{-1}\Omega^{2}wn^{a} + \Omega D^{a}w$ satisfies the conditions $\Lie_{\psi}n^{a}~\hat{=}~0$ and $\Lie_{\psi}\qmet^{a b}~\hat{=}~0$.
	 \end{enumerate}
	\end{theorem}
The proof of the above theorem is fairly straightforward and appears in~\ref{proof:ASAM:identify}. We can also hardly overemphasize the role of linearity of limits and Lie derivatives in arriving at the results obtained so far can. As expected (and also shown in Theorem~\ref{thm:linearity} in~\ref{append:linearity}) the asymptotic symmetry generators inherit linearity as a consequence of their defining properties. In view of this theorem, asymptotic symmetry generators strongly resemble Killing vectors.

The above analysis was primarily concerned about the mathematical properties of asymptotic symmetry generators. In particular, in Definition~\ref{def:ASAM} all kinds of asymptotic symmetry generators are treated on an equal footing. However, the analysis leading to Theorem~\ref{thm:ASAM:props} shows that there are two very distinct kinds of generators (which have been denoted by $\tau^a$ and $\psi^a$ above), distinguished both by their asymptotic profile as well as by the way they act on the ripples. These observations, unsurprisingly, are directly linked to the physical interpretation of these two kinds of generators.

Let us begin with the generator $\tau^{a}$ given in equation~\Eqref{def:tau^a}, whose action on unit hyperboloid completion functions are determined using equation~\Eqref{expr:Ltau_O}. Now, define the function $\Omega'$ as follows
	\begin{eqnarray*} 
	 \Omega'&~\equiv~\Omega + \Lie_{\tau}\Omega \\
	        &~=~(1 + w\Omega)\Omega~.
	\end{eqnarray*}
Clearly, the function $\Omega'$ is just another unit hyperboloid completion function, related to  $\Omega$ via an equation of the form~\Eqref{def:Omega'} with $\alpha = 1 + w\Omega$. This implies that the vector field $\tau^{a}$ generates transformations between unit hyperboloid functions. Consequently, the function $w$ is~\emph{exactly analogous} to the function $\omega$. Indeed, as the above example shows, the correspondence holds beyond the superficial resemblance between the respective definitions~\Eqref{def:w} and~\Eqref{def:omega}.

However, if the function $w$ were only to be restricted by the conditions given in Theorem~\ref{thm:w}, then the mass aspect associated with the completion function $\Omega'$ defined above could potentially be singular on $\scrH$ based on our previous discussion. Clearly, such functions cannot be included into the set of physically viable completion functions. In other words, consistency demands that the function $w$ be given by
	\begin{equation}\label{expr:w}
	 w = \vartheta\log\Omega~, \qquad\qquad \vartheta\,\in\,\bTheta~,
	\end{equation}
exactly as how the function $\omega$ appears in equation~\Eqref{expr:omega}. As a result, $w$ will also satisfy
	\begin{equation}\label{w':sing}
	 \lim_{\Omega \to 0}\Omega\frac{\dl w}{\dl\Omega} \neq \infty~,
	\end{equation}
just like equation~\Eqref{omega':sing}, which is strictly stronger than the analogous condition on the `$\Omega$-derivative' of $w$ appearing in equation~\Eqref{w:sing}. Note, from~\Eqref{Ltau_dO:H} --~\Eqref{Ltau_qmet^ab:H}, that the $\tau^{a}$-generators have a trivial action on $\scrH$ since they vanish asymptotically. Henceforth, we will call asymptotic symmetry generators like $\tau^{a}$ as~\emph{generators of supertranslations}, anticipating their role as the generators of the Lie algebra $\str$ associated with the group $\Str$ defined in~\Eqref{def:Str}. We will justify this identification in the following section.

Let us now consider the $\psi^{a}$-generators. They are tangential to the $\Omega =$ constant hypersurfaces everywhere according to~\Eqref{expr:Lpsi_O}, and preserve the vector $n^{a}$ asymptotically as shown in~\Eqref{Lpsi_n^a:H} and hence do not affect the ripples. On the other hand, from~\Eqref{KVE:psi:H}~, these generators admit a non-zero asymptotic limit and become Killing vectors of the unit hyperboloid metric on $\scrH$. The algebra of Killing vectors of the unit hyperboloid metric is well known to be isomorphic to the algebra of Lorentz transformations. We will show below that indeed the $\psi^{a}$-generators become linear combinations of Lorentz symmetry generators asymptotically. Hence we will denote such generators as~\emph{asymptotic Lorentz generators}.

Of course, there will be a subset of generators which share the features of both the above kinds, namely those which have a trivial effect on the ripples and also vanish on $\scrH$. Recall that the asymptotic symmetry generators which act as identity transformation on the universal structure are~\emph{gauge generators}. Therefore, a gauge generator may only relate unit hyperboloid completion functions within a given ripple, and it cannot approach a non-trivial Killing vector of the unit hyperboloid metric on $\scrH$ in the asymptotic limit. From equation~\Eqref{def:w}, we can see that a generator $\xi^{a}$ will have a trivial effect on the ripples if and only if the function $w$ vanishes asymptotically. Therefore, if an asymptotic symmetry generator satisfies $w~\hat{=}~0$ and vanishes asymptotically, then it is necessarily a gauge generator.

Na\"ively, it may appear that a $\scrH$-preserving vector field $X^{a}$ which vanishes asymptotically, and for which the function $\Omega^{-2}\Lie_{X}\Omega$ also does the same, might qualify as a gauge generator. However, the vector $\taup^{a}$ introduced in equation~\Eqref{def:taup^a}, for any non-trivial choice of the function $w$, provides a counterexample to dispel such a claim. This motivates the need to carefully formulate the criteria for gauge generators in the spirit of Theorem~\ref{thm:ASAM:identify}.

With that as our goal, let $\Lambda^{a}$ be a $\scrH$-preserving vector field. The preceding observations, including the relation~\Eqref{expr:w}, suggest that a necessary (but not sufficient) condition for $\Lambda^{a}$ to have trivial effect on the ripples is to require
	\begin{equation}\label{def:vth0}
	 \Lambda^{a}\Dl_{a}\Omega = \Omega^{2}\log\Omega\,\vartheta_{0}~, \qquad\qquad \vartheta_{0}\,\in\,\bThetao~.
	\end{equation}
An orthogonal decomposition of $\Lambda^{a}$ with respect to the $\Omega =$ constant hypersurfaces then takes the form
	\begin{equation*}
	 \Lambda^{a} = \Lan^{a} + \Lap^{a}~,
	\end{equation*}
where we have have defined
	\begin{equation}\label{def:LanLap^a}
	 \Lan^{a}~\equiv~\frac{(\Omega^{2}\log\Omega)\vartheta_{0}}{F}n^{a}~, \qquad\qquad\qquad \Lap^{a}~\equiv~\tensor{\pmet}{^{a}_{b}}\Lambda^{b}~.
	\end{equation}
Both $\Lan^{a}$ and $\Lap^{a}$ are $\scrH$-preserving vector fields by construction. Using~\Eqref{Ltaup_dO:H} --~\Eqref{Ltaup_qmet^ab:H} we can easily show that $\Lan^{a}$ is in fact a gauge generator. Therefore, from Theorem~\ref{thm:ASAM:identify}, to show that $\Lambda^{a}$ is  a gauge generator, we need to further establish
	\begin{equation}\label{Lap:gauge}
	 \Lap^{a}~\hat{=}~0~, \qquad\qquad \Lie_{\Lap}n^{a}~\hat{=}~0~, \qquad\qquad \Lie_{\Lap}\qmet^{a b}~\hat{=}~0~.
	\end{equation}
Now, with the help of the results used in the proof of Theorem~\ref{thm:HPVF}, we can establish that the most general form of the vector $\Lap^{a}$ satisfying the above conditions is 
	\begin{equation}\label{expr:Lap^a}
	 \Lap^{a} = \lambda_{1}L_{1}^{a} + \lambda_{2}L_{2}^{a} + \cdots~,
	\end{equation}
where
	\begin{itemize}
	 \item each $L_{\scI}^{a}$ (for $\scI = 1,\,2,\,\cdots$) is a $\scrH$-preserving vector field which is tangential to the $\Omega =$ constant hypersurfaces but otherwise completely general (\ie, neither preserves $n^{a}$ nor $\qmet^{a b}$ asymptotically);
	 \item the functions $\lambda_{\scI}$ (for every value of the index $\scI$) are at least $C^{1}$ in $\nhood$ as well as on $\scrH$, and necessarily satisfy
	\end{itemize}
	\begin{equation}\label{lambda:gauge}
	 \lim_{\Omega \to 0}\lambda_{\scI} = 0~;
	\end{equation}
	\begin{itemize}
	 \item at least one  of the following is true:
	  \begin{itemize}
	   \item every single one of the vectors $L_{\scI}^{a}$ vanish asymptotically;
	   \item if the asymptotic limit of one of the $L_{\scI}^{a}$-vectors is non-zero, then the corresponding function $\lambda_{\scI}$ satisfies
	  \end{itemize}
	\end{itemize}
     	\begin{equation}\label{lambda':gauge}
	 \lim_{\Omega \to 0}\frac{\dl\lambda_{\scI}}{\dl\Omega} = 0~,
	\end{equation}
in addition to the condition~\Eqref{lambda:gauge}. Putting all these information together, the most general form of a gauge generator is thus
	\begin{equation}\label{expr:Lambda^a}
	 \Lambda^{a} = \frac{(\Omega^{2}\log\Omega)\vartheta_{0}}{F}n^{a} + \lambda_{1}L_{1}^{a} + \lambda_{2}L_{2}^{a} + \cdots~.
	\end{equation}
It should be obvious that the various conditions defining the functions $\vartheta_{0}$, $\lambda_{1}$, $\lambda_{2},\,\cdots$ mentioned above are in fact completion independent.

We should also note that while Definition~\ref{def:ASAM} explicitly refers to some unit hyperboloid completion to prescribe the defining properties of asymptotic symmetry generators, nevertheless:
	\begin{itemize}
	 \item an asymptotic symmetry generator with respect an unit hyperboloid completion remains so with respect to~\emph{every} unit hyperboloid completion;
	 \item a gauge generator with respect an unit hyperboloid completion cannot generate physical transformations under a change of completion (\ie, it remains a gauge generator with respect to every completion).
	\end{itemize}
The first observation, taken along with Theorem~\ref{thm:ASAM:props}, then implies that every asymptotic symmetry generator will admit a decomposition analogous to~\Eqref{xi=tau+psi} with respect to every completion. However the part of the generator to be regarded as a supertranslations generator and/or an asymptotic Lorentz generator depends on the completion itself. To eleborate, let $\xi^{a}$ be an asymptotic symmetry generator and let $\tau^{a}$ and $\psi^{a}$ be, respectively, the associated supertranslations and asymptotic Lorentz generator parts with respect to the completion $(\hat{\man}, \Omega)$ as per equation~\Eqref{xi=tau+psi}. Then with respect to another such completion $(\hat{\man}, \Omega' = (1 + \omega\Omega)\Omega)$, we have the decomposition
	\begin{equation}\label{xi=tau'+psi'}
	 \xi^{a} = \tau^{\prime a} + \psi^{\prime a}~,
	\end{equation}
where $\tau^{\prime a}$ represents the supertranslations generating part of $\xi^{a}$ with respect to $(\hat{\man}, \Omega')$, and $\psi^{\prime a}$ denotes the corresponding asymptotic Lorentz generating part. In particular $\psi^{\prime a}\Dl_{a}\Omega' = 0$. The generators $\tau^{\prime a}$ and $\psi^{\prime a}$ themselves are given by
	\begin{equation}\label{def:tau'_psi'}
	 \tau^{\prime a}~\asymp~\tau^{a} + \tau_{\Lie_{\psi}\omega}^{a}~, \qquad\qquad\qquad \psi^{\prime a}~\asymp~\psi^{a} - \tau_{\Lie_{\psi}\omega}^{a}~,
	\end{equation}
where the symbol `$\asymp$' denotes~\emph{equality up to a gauge generator} between asymptotic symmetry generators, such that for two generators $\xi_{1}^{a}$ and $\xi_{2}^{a}$
	\begin{equation}\label{def:asymp}
	 \xi_{1}^{a}~\asymp~\xi_{2}^{a} \qquad\qquad\iim\qquad\qquad \xi_{1}^{a} = \xi_{2}^{a} + \Lambda^{a}~, 
	\end{equation}
for some gauge generator $\Lambda^{a}$\footnote{Note that $\Lambda^{a} \asymp 0$ in the above notation. It is trivial to verify that `$\asymp$' describes an equivalence relation. Indeed, the three equivalence relations introduced so far -- namely, `gauge equivalence' between unit hyperboloid completion functions, `asymptotic equivalence' between the functions $\Theta \in \bTheta$ and `equality up to a gauge generator' between asymptotic symmetry generators introduced above -- are manifestations of the same gauge redundancy associated with the asymptotic symmetry transformations.}. Finally, the subscript on the supertranslations generator $\tau_{\Lie_{\psi}\omega}^{a}$ with respect to $(\hat{\man}, \Omega)$ denotes the `$w$-function' associated with it, in accordance with equation~\Eqref{def:tau^a}.

The above relations also show that a~\emph{pure} supertranslations generator remains invariant (up to a gauge generator) under a change of completions. This is clearly not true for asymptotic Lorentz generators, since such generators cannot be tangential to the $\Omega =$ constant surfaces in every completion. Rather, a pure asymptotic Lorentz generator in one completion appears as a general linear combination of both types of generators (up to a gauge generator) in another completion. However since supertranslation generators vanish on $\scrH$, the asymptotic Lorentz generating parts of a given symmetry generator with respect to two completions admit the same asymptotic limit. In particular, if we introduce the following notation
	\begin{equation}\label{def:upsi}
	 \upsi^{a}~\equiv~\lim_{\Omega \to 0}\psi^{a}~,
	\end{equation}
for the asymptotic limit of the generator $\psi^{a}$ in the completion $(\hat{\man}, \Omega)$, then 
	\begin{equation}\label{psi_prime:H}
	 \lim_{\Omega \to 0}\psi^{\prime a}~=~\upsi^{a}~,
	\end{equation}
with respect to the completion $(\hat{\man}, \Omega')$.

This concludes our discussion of the basic properties of asymptotic symmetry generators. In the following section we will study the algebra of these generators, and in the process justify the terms supertranslations generators as well as asymptotic Lorentz generators.
\subsection{Asymptotic symmetry generators: algebra}\label{sec:ASAM:algebra}
So far, we have obtained a set vector fields called the asymptotic symmetry generators which describe all possible transformations under which the rigid substructure of the universal structure at spatial infinity remains invariant. The Lie algebra $\spi$, by definition, is generated by these vector fields and underlie the Lie group $\Spi$ introduced previously. In the following we will explore the structure of $\spi$.

We will begin with the commutator between two supertranslations generators $\tau_{1}^{a}$ and $\tau_{2}^{a}$ (defined as in equation~\Eqref{def:tau^a} in terms of their respective `$w$-functions' $w_{1}$ and $w_{2}$). A straightforward calculation reveals
	\begin{equation*}
	 [\tau_{1}, \tau_{2}]^{a}~\asymp~\check{\tau}^{a}~,
	\end{equation*}
where the vector field $\check{\tau}^{a}$ also has the form of a supertranslations generator with the associated `$w$-function' being given by
	\begin{equation*}
	 \check{w} = \Omega^{2}\left[w_{1}\frac{\dl w_{2}}{\dl\Omega} - w_{2}\frac{\dl w_{1}}{\dl\Omega}\right].
	\end{equation*}
Note that if $w_{1}$ and $w_{2}$ are only restricted by the weaker smoothness conditions of Theorem~\ref{thm:w}, then $\check{\tau}^{a}$ is yet another non-trivial supertranslations generator. This seems to indicate that two supertranslations generators need not commute in general, unlike the corresponding result for smooth `$w$-functions'~\cite{Ashtekar:1991vb}.

This is exactly analogous to the situation with finite transformations between completion functions encountered earlier: as we saw, two such transformations will not commute unless the corresponding `$\omega$-functions' are further restricted to ensure finiteness of the mass aspect with respect to every unit hyperboloid completion. Here too, we see~\emph{exactly} the same story: the generator $\check{\tau}^{a}$ becomes a gauge generator under the stronger smoothness requirement~\Eqref{w':sing} for the functions $w_{1}$ and $w_{2}$, resulting in asymptotic commutativity of the supertranslations generators, \ie
	\begin{equation}\label{spi:[tau,tau]}
	 [\tau_{1}, \tau_{2}]^{a}~\asymp~0~.
	\end{equation}
The generators of supertranslations span the Lie algebra $\str$ associated with the group of supertranslations $\Str$.

The analysis of the asymptotic Lorentz generators is considerably simpler. In particular, if $\psi_{1}^{a}$ and $\psi_{2}^{a}$ are two such generators, then their commutator $[\psi_{1}, \psi_{2}]^{a}$ can easily be seen to also be tangential to the $\Omega =$ constant hypersurfaces such that
	\begin{equation*}
	 [\psi_{1}, \psi_{2}]^{a}~\hat{=}~\left[\upsi_{1}, \upsi_{2}\right]^{a},
	\end{equation*}
where the right hand side denotes the commutator (evaluated on $\scrH$) of the asymptotic limits of the corresponding generators, employing the notation defined in~\Eqref{def:upsi}. Using Theorem~\ref{thm:ASAM:identify}, we can show that the commutator $[\psi_{1}, \psi_{2}]^{a}$ is an asymptotic symmetry generator. Hence, the set of all such commutators must be isomorphic to the Lie algebra of the Killing symmetries of the unit hyperbolic metric induced on $\scrH$, \ie,
	\begin{equation}\label{spi:[psi,psi]}
	 \left[\upsi_{1}, \upsi_{2}\right]^{a}~\in~\so(1, 3)~,
	\end{equation}
where $\so(1, 3)$ is the Lie algebra of the Lorentz generators of $(1 + 3)$-dimensional Minkowski spacetime.

Moreover, the commutator between an asymptotic Lorentz generator $\psi^{a}$ and a supertranslations generator $\tau^{a}$ (the latter being expressed in terms of the function $w$ as usual~\Eqref{def:tau^a}) works out to be
	\begin{equation}\label{spi:[psi,tau]}
	 [\psi, \tau]^{a}~\asymp~\tau_{\Lie_{\psi}w}^{a}~,
	\end{equation}
where $\tau_{\Lie_{\psi}w}^{a}$ denotes a supertranslations generator, and $\Lie_{\psi}w$ the associated `$w$-function'. This result is consistent with our earlier finding in equation~\Eqref{def:tau'_psi'} and shows that the actions of supertranslations and asymptotic Lorentz transformations on the ripples do not commute.

Finally, the commutator between any asymptotic symmetry generator and a gauge generator is itself a gauge generator as consistency would demand.

To summarize: the Lie algebra $\spi$ is generated by the set of all asymptotic symmetry generators, with the basic commutation relation being given by
	\begin{equation}\label{spi:[xi,xi]}
	 [\xi_{1}, \xi_{2}]^{a}~\asymp~\tilde{\tau}^{a} + [\psi_{1}, \psi_{2}]^{a}~, \qquad\qquad \tilde{w}~\equiv~\Lie_{\psi_{1}}w_{2} - \Lie_{\psi_{2}}w_{1}~,
	\end{equation}
where
	\begin{itemize}
	 \item the functions $w_{1}$ and $w_{2}$ are the `$w$-functions' associated with the generators $\xi_{1}^{a}$ and $\xi_{2}^{a}$ respectively as per the definition~\Eqref{def:w};
	 \item $\psi_{1}^{a}$ and $\psi_{2}^{a}$ are the respective asymptotic Lorentz generators as per the decomposition~\Eqref{xi=tau+psi};
	 \item $\tilde{\tau}^{a}$ is a supertranslations generator expressed in terms of the function $\tilde{w}$ according to equation~\Eqref{def:tau^a}.
	\end{itemize}
Based on he above commutation relation, we may further conclude that
	\begin{itemize}
	 \item $\str$ and $\so(1, 3)$ are proper subalgebras of $\spi$;
	 \item $\str$ is an ideal of $\spi$;
	 \item $\so(1, 3)$ is not an ideal of $\spi$.
	\end{itemize}
We can therefore express $\spi$ as a  semi-direct sum of the subalgebras $\str$ and $\so(1, 3)$ as follows
	\begin{equation}\label{spi=str+so}
	 \spi = \str \oplus_{\textsc{s}} \so(1, 3)~.
	\end{equation}    
Consequently, the corresponding relation between the Lie groups, with the Lie group $\Str$ being a~\emph{normal subgroup} of $\Spi$, is
	\begin{equation}\label{Spi=Str*OO}
	 \Spi = \Str \rtimes \Lorentz(1, 3)~,
	\end{equation}
where $\Lorentz(1, 3)$ is the group of rotations and boosts in $(1 + 3)$-dimensions.

We conclude this section with a justification of the name `supertranslations'. We start with the observation that the Lie algebra $\str$ admits an abelian subalgebra $\trn$ consisting of the generators
	\begin{equation}\label{def:P_mu^a}
	 P_{\mu}^{a}~\equiv~\frac{\Omega^{2}\varkappa^{\mu}}{F}n^{a} - \Omega D^{a}\varkappa^{\mu}~, \qquad\qquad \mu = 0,\,1,\,2,\,3~,
	\end{equation}
where $\{\varkappa^{\mu}\}_{\mu = 0}^{3}$ are the first four~\emph{hyperbolic harmonics}, \ie, solutions of the equation (see, \eg, reference~\cite{Ashtekar:1991vb} for further details)
	\begin{equation}\label{def:varkappa}
	 D_{a}D_{b}\varkappa^{\mu} + \varkappa^{\mu}\hmet_{a b}~\hat{=}~0~, \qquad\qquad  \mu = 0,\,1,\,2,\,3~.
	\end{equation}
We can verify that the generators in~\Eqref{def:P_mu^a} are the four (spi)~\emph{translation generators} by checking, for example, that their commutation relations with the elements of $\so(1, 3)$ (\eg, based on the general result~\Eqref{spi:[psi,tau]}) reproduce the relevant parts of the Poincar\'e algebra.

Another way to arrive at the same conclusion is to determine the action of these generators on a set of~\emph{asymptotically Cartesian/Minkowskian coordinates}. Such coordinates, by design, emulate the behavior of standard Minkowskian coordinates (of the global Minkowski spacetime) near $\scrH$. A set of such coordinate functions can be represented in terms of a completion function $\Omega$ as follows (see, \eg, references~\cite{ashtekar1985logarithmic, Ashtekar:2008jw})
	 \begin{equation}\label{def:x^mu}
	  x^{\mu} = \frac{\varkappa^{\mu}}{\Omega} + x_{(1)}^{\mu}\log\Omega + x_{(2)}^{\mu}~, \qquad\qquad \mu = 0,\,1,\,2,\,3~,
	\end{equation}
where the leading order term on the right follows from the representation of Minkowskian coordinates in terms of $\Omega$ and the hyperbolic harmonics, while terms involving the functions $\{x_{(1)}^{\mu}\}_{\mu = 0}^{3}$ and $\{x_{(2)}^{\mu}\}_{\mu = 0}^{3}$ (which are $C^{1}$ in $\nhood$ as well as on $\scrH$ by assumption) describe the `departure' of these coordinates from their global Minkowskian nature\footnote{The necessity of including the term involving the functions $x_{(1)}^{\mu}$ stems from the generalized form of the transformation~\Eqref{expr:omega}. In particular, even if one starts with a choice of asymptotically Cartesian coordinates for which $x_{(1)}^{\mu} = 0$ with respect to a given completion, such terms will be generated under a change of completion.}. The action of a general supertranslations generator~\Eqref{def:tau^a} on $x^{\mu}$ then works out to be
	\begin{equation}\label{expr:Ltau_x^mu}
	 \Lie_{\tau}x^{\mu} = C^{\mu}\log\Omega - (D^{c}w)(D_{c}\varkappa^{\mu}) + \cdots~,
	\end{equation}
where
	\begin{equation}\label{def:C^mu}
	 C^{\mu}~\equiv~-\vartheta\varkappa^{\mu}~, \qquad\qquad \vartheta\,\in\,\bTheta~,
	\end{equation}
and the `$\cdots$' represent terms which explicitly vanish in the asymptotic limit. Let us emphasize that the expression on the right hand side of~\Eqref{expr:Ltau_x^mu} is~\emph{not} a perturbative expansion in any sense. Rather, in order to highlight the physically relevant parts of the expression, we have chosen to explicitly display only those terms which are either divergent or finite and non-zero in the asymptotic limit. In particular, the functions $\vartheta$ (and hence $C^{\mu}$) as well as $w$ both depend on $\Omega$ in general.

If $\tau^{a}$ is one of the translations generators introduced in~\Eqref{def:P_mu^a}, the right hand side of~\Eqref{expr:Ltau_x^mu} can be easily seen to reduce to either zero or $\pm 1$ (in the asymptotic limit) depending on the choices of the function $x^{\mu}$ as well as of the translations generator. This confirms that the generators $P_{\mu}^{a}$ indeed act as generators of rigid translations near spatial infinity. A more general supertranslations generator then generates `spacetime dependent' translations parametrized by the function $w$. This is what justifies their naming.

Furthermore, for general supertranslations, allowing the function $w$ to be asymptotically singular (but restricted by~\Eqref{expr:w}) allows for the~\emph{logarithmic translations} of the Cartesian coordinates appearing on the right hand side of the result~\Eqref{expr:Ltau_x^mu}. Note that these are the most general kind of such translations since the functions $C^{\mu}$ are spacetime dependent in general. Such generalizations of supertranslations have already been proposed in earlier works 
like~\cite{ashtekar1985logarithmic, Ashtekar:2008jw, Beig:1982ifu, Fuentealba:2022xsz}. However, to reemphasize, we have explicitly linked the origin of the logarithmic supertranslations to the finiteness of the mass aspect. To the best of our knowledge, this has not been demonstrated in the literature before.
\section{Outlook}\label{sec:outlook}
Our description of asymptotic symmetries presented above has been completely geometrical. The basic idea was to call a transformation an asymptotic symmetry transformation if it asymptotically preserved the geometrical structures required by Definitions~\ref{def:AR:asymp} -~\ref{def:AR:AM}. In this sense, we may regard our treatment of asymptotic symmetries as~\emph{kinematical}.

In physics however, we are more accustomed to a~\emph{dynamical} view of symmetries: we associate the concept with transformations which map physical solutions of the equations of motion of some physical system to other physical solutions. Since the asymptotic symmetry generators must have a physical interpretation, they should connect distinct  asymptotically flat (in the sense of Definitions~\ref{def:AR:asymp} -~\ref{def:AR:AM}) solutions of the Einstein's field equations.

More specifically, let $\met_{a b}$ be an asymptotically Minkowskian solution of the Einstein's field equations and let $\tau^{a}$ be a generator of supertranslations. The (linearized) change in the metric generated by $\tau^{a}$ and valid `near $\scrH$' can then be expressed as
	\begin{equation}\label{def:delta[tau]_met}
	 \delta\met_{a b}~\equiv~\epsilon\Lie_{\tau}\met_{a b}~, \qquad\qquad 0 < \epsilon \ll 1~,
	\end{equation}
where $\epsilon$ is a constant book-keeping parameter quantifying parametric smallness of the transformation with respect to the initial solution.

An orthogonal decomposition of the transformed solution $(\met_{a b} + \delta\met_{a b})$, with respect to the $\Omega =$ constant hypersurfaces, reads
	\begin{equation}\label{def:perturbed_met}
	 \met_{a b} + \delta\met_{a b} = \frac{1 + 2\epsilon\Omega\pertS}{\Omega^{4}F}\Dl_{a}\Omega\Dl_{b}\Omega + \frac{\epsilon\Omega(\pertV_{a}\Dl_{b}\Omega + \pertV_{b}\Dl_{a}\Omega)}{\Omega^{2}} + \frac{\qmet_{a b} + 2\epsilon\Omega\pertT_{a b}}{\Omega^{2}}~,
	\end{equation}
where the scalar $\pertS$, the one-form $\pertV_{a}$ and the rank-$(0, 2)$ symmetric tensor $\pertT_{a b}$ are given by
	\begin{eqnarray}\label{def:pertSVT}
	 & \pertS = \frac{\mu_{0}}{2}~, \\
	 & \pertV_{a} = -D_{a}\left[\frac{\dl w}{\dl\Omega}\right] + \qmet^{(1)}_{a c}D^{c}w - \mu D_{a}w~, \\
	 & \pertT_{a b} = -w\hmet_{a b} - D_{a}D_{b}w~,
	\end{eqnarray}
where
	\begin{itemize}
	 \item $\mu_{0}$ is the constant associated with the function $w$ according to equation~\Eqref{omega':sing} and $\mu$ is the mass aspect with respect to the completion $(\hat{\man}, \Omega)$;
	 \item $\hmet_{a b}$ is the standard unit hyperboloid metric in $(1 +2)$-dimensions, while
	  \begin{equation}\label{def:q1met_ab}
	   \qmet^{(1)}_{a b}~\equiv~\frac{\qmet_{a b} - \hmet_{a b}}{\Omega}~,
	  \end{equation}
such that $\Omega\qmet^{(1)}_{a b}$ vanishes asymptotically;
	 \item $\pertV_{a}$ and $\pertT_{a b}$ are both tangential to the $\Omega =$ constant hypersurfaces (in the sense $\pertV_{a}n^{a} = 0$ and $\pertT_{a b}n^{b} = 0$).
	\end{itemize}
Note that while the expression on the right hand side of~\Eqref{def:perturbed_met} represents an approximation of the transformed solution, which becomes increasingly accurate as one goes to successively smaller values of $\Omega$, it must~\emph{not} be interpreted as `a perturbative expansion in $\Omega$ around $\Omega = 0$' (\eg, in the sense of~\cite{Beig:1982ifu}). Rather, the functions $\pertV_{a}$ and $\pertT_{a b}$ are both dependent to $\Omega$ by construction. Indeed, both $\pertV_{a}$ and $\pertT_{a b}$ are potentially divergent in the asymptotic limit\footnote{Note, in particular, that the first term in $\pertT_{a b}$ diverges like $\log\Omega$ in the asymptotic limit, owing to the relation~\Eqref{expr:w}. This is strong evidence that the transformation considered describes a logarithmic supertranslation.}, but their contributions in the transformed solution vanish in the same limit owing to the factors of $\Omega$ appearing with each term. This is also why, we have neglected all other terms in their expressions (as well as in that of $\pertS$) which vanish in the asymptotic limit. Nevertheless, in the regime where our approximation is consistent with those in earlier work (see, \eg,~\cite{Compere:2011ve, Troessaert:2017jcm}) there seems to be agreement.

That being said, to fully explore the consequences of the generalizations proposed in this work, a more detailed~\emph{non-perturbative} (\ie, beyond the above-mentioned linearization) analysis of the effect of the symmetry generators on the space of asymptotically Minkowskian solutions is highly desirable. It is important to stress in this regard that unlike the case of `smooth spacetimes' studied by~\cite{Beig:1982ifu, Beig:1983sw, Ashtekar:1991vb}, we do not have the luxury of performing `series expansions around $\Omega = 0$' to approximate the behavior of a general asymptotically Minkowskian spacetime to some desired accuracy. Rather, here we require a full-fledged study of the problem of~\emph{radial evolution} -- \ie, how to relate the data on one $\Omega =$ constant surface to another one (at least in $\nhood$, if not globally) via the Einstein's equations. Naturally, issues such as regularity of the `initial data' (\ie, that which is prescribed on an `initial' $\Omega =$ constant hypersurface) as well as the stability of the solutions under changes to that data (through asymptotic symmetry transformations) become very important in such contexts. In particular, since the aforementioned radial evolution takes place in a spacelike direction, questions of stability have to be carefully analyzed to check if it further restricts the asymptotic transformations (\ie, the function $\Theta$). Understandably, problems like these rightfully deserve rigorous and extensive investigations which was beyond the scope of the current project. Incidentally, studies on the minimal smoothness allowed in the initial data which still guarantees the existence of a regular mass aspect and angular momentum aspect at null infinity have been previously carried out, \eg, in~\cite{Andersson:1994ng, Friedrich:1998xmu, Bieri:2009xc, Kehrberger2024}. However, whether these results agree with our generalizations remains to be understood.

Another interesting question is how the enlarged class of asymptotic symmetry transformations proposed in this work affect the behavior of specific solutions of the Einstein's equations.~\ref{append:Schwarzschild_UHC} attempts to illustrate how may one construct a unit hyperbolic completion of the Schwarzschild spacetime. However, a detailed analysis of such examples could be better appreciated and contextualized once the physical significances of the enlarged class of asymptotic symmetry transformations are better understood; we wish to present them in our follow up work.

\color{black}

Additionally, although our analysis is completely covariant, it is essential to compare and contrast our results with those obtained in a specific gauge like in~\cite{Henneaux:2018cst, Henneaux:2018gfi, Henneaux:2018hdj, Henneaux:2019yax, Prabhu:2019daz} among others. Such comparisons may reveal the need for further restrictions on the function $\Theta$ beyond those listed in equations~\Eqref{Theta:sing} and~\Eqref{Theta'':sing}.

It will also be very interesting to understand the connection between the requirement of finiteness of the mass aspect with respect to every completion (as proposed in the current work) with the requirement of finiteness of the action of the underlying theory as has been proposed recently in~\cite{Henneaux:2019yax, Fuentealba:2022xsz}. Other important questions we wish to address in the future include the derivation of the Noether charges associated with asymptotic symmetries. For smooth completions, this has already been attempted in the past (see, \eg,~\cite{Ashtekar:1990gc, Perng:1998vn}) and it will be interesting to verify how these results generalize once we take into account the broader class of supertranslations proposed in this work. Another interesting challenge would be to investigate whether it is possible to understand the relations between the asymptotic symmetry groups at different (\ie, timelike, null and spatial) infinities of asymptotically flat spacetimes within the current framework. Our formalism, with suitable modifications, can also be used to understand asymptotic behavior of spacetimes which do not asymptote to Minkowski but to some other maximally symmetric spacetimes like de-Sitter or anti de-Sitter.
\section*{Acknowledgment}
SM acknowledges the financial support of Council for Scientific and Industrial Research (CSIR), India under grant number 09/0919(11323)/2021-EMR-I. Discussions with Shankhadeep Chakrabortty is gratefully acknowledged.The authors extend their sincere thanks to Alok Laddha for his expert review and constructive criticisms. Some preliminary calculations leading to some of the presented results, as well as the analysis presented in~\ref{append:Schwarzschild_UHC} were performed using the~\texttt{SageManifolds} functionalities of the free and open-source computer algebra system~\texttt{SageMath}~\cite{sagemath}.
\appendix
\section{Proofs etc.}\label{append:proofs}
\subsection{Proof of Theorem~\ref{thm:omega}}\label{proof:omega}
\begin{proof}
The proof of the first part of the theorem is essentially given in the discussion preceding the statement of the theorem.

For the second part of the theorem, the case of $\omega$ being finite on $\scrH$ need not require any special attention. Therefore, let us only focus on the case where $\omega$~\emph{does} diverge asymptotically but in a manner restricted by the requirements of the theorem.

Now, since $\omega$ diverges asymptotically but $D_{a}\omega$ stays finite in the same limit, we must have
	\begin{equation*}
	 D_{a}\log\omega~\hat{=}~0~.
	\end{equation*}
Na\"ively, this should mean that $\log\omega$ can be expressed as the sum of a constant and a function which vanishes asymptotically (just like the functions $F$ or $\alpha$; recall our earlier results~\Eqref{F=1:H} -~\Eqref{def:alpha}). However, $\log\omega$ must also diverge asymptotically since $\omega$ does so by assumption. These two fact can be simultaneously true~\emph{only if} the divergence in $\log\omega$ comes from a function of just $\Omega$ which appears additively in $\log\omega$. Equivalently, the divergent part of $\omega$ must be a function of just $\Omega$ and must appear multiplicatively in $\omega$.

Let $\sigma_{0}$ be a\footnote{Note that $\sigma_{0}$ is not unique; rather, given a choice for $\sigma_{0}$ another valid choice would be $f\sigma_{0}$ for any function $f$ of $\Omega$ which approaches the value one asymptotically.} divergent function of just $\Omega$ describing the dominant asymptotic behavior of $\omega$. Then, the function
	\begin{equation*}
	 \wp_{0}~\equiv~\sigma_{0}^{-1}\omega~,
	\end{equation*}
must be~\emph{finite} and~\emph{non-vanishing} asymptotically. Indeed, the fact that we have~\emph{correctly} identified the dominant asymptotic behavior of $\omega$ must mean
	\begin{equation*}
	 \lim_{\Omega \to 0}\wp_{0} = 1~.
	\end{equation*}
Now, in order for $\omega$ to respect the conditions in~\Eqref{omega:sing}, the function $\sigma_{0}$ must obey the criteria~\Eqref{sigma:sing} and~\Eqref{sigma':sing}. However, since $D_{a}\wp_{0} = \sigma_{0}^{-1}D_{a}\omega$, we must also have
	\begin{equation*}
	 D_{a}\wp_{0}~\hat{=}~0~.
	\end{equation*}
Since $\wp_{0}$ approaches the value one in the asymptotic limit as observed before, the above must mean
	\begin{equation*}
	 \wp_{0} = 1 + \tilde{\omega}_{0}~,
	\end{equation*}
where the function $\tilde{\omega}_{0}$ is asymptotically vanishing. We may now express $\tilde{\omega}_{0}$ as
	\begin{equation*}
	 \tilde{\omega}_{0} = \frac{\omega_{1}}{\sigma_{0}}~,
	\end{equation*}
such that $\omega_{1}$~\emph{may or may not} diverge on $\scrH$. If $\omega_{1}$ assumes a finite asymptotic limit, then $\tilde{\omega}_{0}$ vanishes on $\scrH$ manifestly. Otherwise, the divergence in $\omega_{1}$ must be~\emph{strictly weaker} than that of $\sigma_{0}$ such that $\tilde{\omega}_{0}$ still vanishes asymptotically. As a consequence of the above observations, we have
	\begin{equation*}
	 \omega  = \sigma_{0}\wp_{0} = \sigma_{0} + \omega_{1}~.
	\end{equation*}
Hence, $\omega_{1}$ must also satisfy the criteria in~\Eqref{omega:sing}. Thus, $\omega_{1}$ must also a~\emph{bona fide} `$\omega$-function'.

If $\omega_{1}$ is asymptotically finite, we identify $\sigma$ with $\sigma_{0}$ and $\omegaf$ with $\omega_{1}$ to arrive at the desired conclusion~\Eqref{o=s+f}. Else, as before, the divergence in $\omega_{1}$ must come from a divergent function of just $\Omega$. Let $\sigma_{1}$ be a divergent function of $\Omega$ only describing the leading order divergent behavior of $\omega_{1}$ asymptotically. Repeating the above arguments, we may then conclude
	\begin{equation*}
	 \omega_{1} = \sigma_{1} + \omega_{2}~, 
	\end{equation*}
where $\sigma_{1}$ again satisfies conditions~\Eqref{sigma:sing} as well as~\Eqref{sigma':sing}, and $\omega_{2}$ is yet again a genuine `$\omega$-function' satisfying all the criteria of the theorem. Hence
	\begin{equation*}
	 \omega = \sigma_{0} + \sigma_{1} + \omega_{2}~.
	\end{equation*}
As before, $\omega_{2}$ could be either asymptotically finite -- whence we have arrived at the desired conclusion~\Eqref{o=s+f} -- or not. In case of the latter possibility, we repeat the above process.

We may keep repeating the process until, after $\scN$ steps ($\scN$ being finite or countably infinite as expected on physical grounds) say, we end up with a function $\omega_{\scN}$ which~\emph{is} finite on $\scrH$. If we now define
	\begin{equation*}
	 \sigma~\equiv~\sum_{\scA\,=\,0}^{\scN}\sigma_{\scA}~,
	\end{equation*}
which clearly satisfies the conditions~\Eqref{sigma:sing} and~\Eqref{sigma':sing}, and
	\begin{equation*}
	 \omegaf~\equiv~\omega_{\scN}~,
	\end{equation*}
then we finally arrive at the desired conclusion~\Eqref{o=s+f}.

In the above proof, the significance of the `linear-in-$\omega$' property of the conditions~\Eqref{omega:sing} is worth stressing.

To establish the last part of the theorem, note that since $\omegaf$ is asymptotically finite, so must be $D_{a}\omegaf$. Furthermore, since $\omega$, $\sigma$ and $\omegaf$ are related linearly~\Eqref{o=s+f}, the condition~\Eqref{sigma':sing} as well as the middle one in equation~\Eqref{omega:sing} imply
	\begin{equation*}
	 \lim_{\Omega \to 0}\Omega^{2}\frac{\dl\omegaf}{\dl\Omega} = 0~.
	\end{equation*}
However, the above is weaker than the middle condition in~\Eqref{omegaf:sing}, and the challenge is to establish the latter. To that end, let
	\begin{equation*}
	 f~\equiv~\Omega^{2}\frac{\dl\omegaf}{\dl\Omega} \qquad\qquad\iim\qquad\qquad \frac{\dl\omegaf}{\dl\Omega} = \frac{f}{\Omega^{2}}~.
	\end{equation*}
Then the preceding observations indicate
	\begin{equation*}
	 \lim_{\Omega \to 0}\,f = 0~.
	\end{equation*}
Upon integrating the above representation of $(\dl\omegaf/\dl\Omega)$ with respect to $\Omega$ between two $\Omega =$ constant hypersurfaces (denoted by $\Omega$ and $\Omega_{0}$, respectively, both of which belong to the neighborhood $\nhood$ by assumption), we obtain
	\begin{equation*}
	 \omegaf|_{\Omega} = \omegaf|_{\Omega_{0}} + \int_{\Omega_{0}}^{\Omega}\frac{\dd\tilde{\Omega}\,f}{\tilde{\Omega}^{2}}~,
	\end{equation*}
where $\tilde{\Omega}$ is a `dummy variable of integration'\footnote{Of course, one needs to set up some suitable coordinate system on the $\Omega =$ constant hypersurfaces to carry out the above integral, and the values of these coordinate functions are held fixed while performing the integral.}. Note that $\omegaf|_{\Omega_{0}}$ is independent of $\Omega$ and only depends on the coordinate functions on the $\Omega = \Omega_{0}$ hypersurface.

The above expression obscures the fact that $\omegaf$ must be finite asymptotically. In fact, let us momentarily relax this requirement by allowing $\omegaf$ to diverge asymptotically,~\emph{but} more mildly than $\log\Omega$. The following limit can then be computed using the L'H\^opital trick
	\begin{equation*}
	 \lim_{\Omega \to 0}\,\frac{\omegaf}{\log\Omega} = \lim_{\Omega \to 0}\,\frac{\Omega^{-2}f}{\Omega^{-1}} = \lim_{\Omega \to 0}\,\Omega^{-1}f~.
	\end{equation*}
The assumed asymptotic nature of $\omegaf$ now necessarily requires the limit on the left hand side to vanish. Consequently, we must also have
	\begin{equation*}
	 \lim_{\Omega \to 0}\,\Omega^{-1}f = 0 \qquad\qquad\iim\qquad\qquad \lim_{\Omega \to 0}\,\Omega\frac{\dl\omegaf}{\dl\Omega} = 0~.
	\end{equation*}
However, since $\omegaf$ is actually asymptotically finite, the above condition must hold for it even more strongly. We have thus established the middle condition in equation~\Eqref{omegaf:sing}. This completes the proof of the theorem.
\end{proof}
\subsection{Three-metrics induced by $\qmet_{a b}$ and $\qmet'_{a b}$ on $\scrH$}\label{proof:same_uhm_on_H}
Let $\{\ehat^{i}_{a}\}_{i = 0}^{2}$ be a basis of one-forms and $\{\fhat_{i}^{a}\}_{i = 0}^{2}$ be an analogous basis of vectors in $\nhood$ as well as on $\scrH$ such that
	\begin{equation*}
	 \ehat^{i}_{a}n^{a} = 0~, \qquad\qquad \fhat_{i}^{a}\Dl_{a}\Omega = 0~, \qquad\qquad \fhat_{i}^{a}\ehat^{j}_{a} = \tensor{\delta}{_{i}^{j}}~.
	\end{equation*}
We stress that the above relations hold~\emph{on} $\scrH$ as well as in $\nhood$. Indeed, to ensure that they hold off $\scrH$, we require that the Lie derivatives of them along $n^{a}$ must vanish. Consequently
	\begin{equation*}
	 \Lie_{n}\ehat^{i}_{a} = \tensor{\calN}{^{i}_{j}}\ehat^{j}_{a}~,
	\end{equation*}
where $\{\tensor{\calN}{^{i}_{j}}\}_{i, j = 0}^{2}$ are nine functions admitting finite limits to $\scrH$. Likewise
	\begin{equation*}
	 \Lie_{n}\fhat_{i}^{a} = -F^{-1}\dot{F}_{i}n^{a} - \tensor{\calN}{^{j}_{i}}\fhat_{j}^{a}~,
	\end{equation*}
where
	\begin{equation*}
	 \dot{F}_{i}~\equiv~\fhat_{i}^{a}\Dl_{a}F = (\fhat_{i}^{a}\Dl_{a}\mu)\Omega~,
	\end{equation*}
where the expression after the second equality follows as a consequence of~\Eqref{def:mu}. Clearly, $\dot{F}_{i}$ vanishes on $\scrH$.

Given a choice of the sets of basis co/vectors on a $\Omega =$ constant ($\neq 0$) hypersurface in $\nhood$, the above Lie derivative relations are capable of propagating the orthogonality relations between the basis vectors and covectors to other such surfaces in $\nhood$. As well, the relations admit finite limits to $\scrH$ (since all such relations are expressed in terms of quantities which admit such limits). In particular the relation for $\Lie_{n}\ehat^{i}_{a}$ retains its form on $\scrH$, while that for $\Lie_{n}\fhat_{i}^{a}$ becomes\footnote{In~\cite{Ashtekar:1991vb} a coordinate basis with coordinate functions $\{\varphi^{i}\}_{i = 0}^{2}$ foliating $\scrH$ was used, which can be transported into the bulk upon solving the equations $\Lie_{n}\varphi^{i} = 0$ (\ie, $n^{a}\Dl_{a}\varphi^{i} = 0$). For such functions, the basis covectors are $\ehat^{i}_{a} = \Dl_{a}\varphi^{i}$, while the basis vectors are $\fhat_{i}^{a} = (\dl_{i})^{a}$. Naturally $\Lie_{n}\ehat^{i}_{a} = 0$ as well as $\Lie_{n}\fhat_{i}^{a} = 0$. These equations are valid on $\scrH$ as well (which is consistent with what we found above).}
	\begin{equation*}
	 \Lie_{n}\fhat_{i}^{a}~\hat{=}~-\tensor{\calN}{^{j}_{i}}\fhat_{j}^{a}~,
	\end{equation*}
Consider, finally, the expansions of the tensors $\qmet_{a b}$ and $\qmet^{a b}$ in these bases
	\begin{equation}\label{expand:qmet}
	 \qmet_{a b} = \qmet_{i j}\ehat^{i}_{a}\ehat^{j}_{b}~, \qquad\qquad \qmet^{a b} = \qmet^{i j}\fhat_{i}^{a}\fhat_{j}^{b}~,
	\end{equation}
where the functions $\qmet_{i j}$ and $\qmet^{i j}$ are defined through
	\begin{equation*}
	 \qmet_{i j}~\equiv~\qmet_{a b}\fhat_{i}^{a}\fhat_{j}^{b}~, \qquad\qquad \qmet^{i j}~\equiv~\qmet^{a b}\ehat^{i}_{a}\ehat^{j}_{b}~,
	\end{equation*}
and are related via
	\begin{equation*}
	 \qmet^{i k}\qmet_{k j} = \tensor{\delta}{^{i}_{j}}~, \qquad\qquad \qmet_{i k}\qmet^{k j} = \tensor{\delta}{_{i}^{j}}~.
	\end{equation*}
The restrictions of the respective coefficients of the above expansions on $\scrH$ will be denoted as follows
	\begin{equation}
	 \qmet_{i j}~\hat{=}~h_{i j}~, \qquad\qquad \qmet^{i j}~\hat{=}~h^{i j}~.
	\end{equation}
The functions $h_{i j}$ are the components of the unit hyperboloid metric in the chosen basis, by the field equations~\cite{Ashtekar:1991vb}. This is the reason the completion is called a~\emph{unit hyperboloid} completion.

For the rest of the discussion, it will suffice to restrict ourselves on the boundary $\scrH$. We may construct another basis $\{\ehat^{\prime i}_{a}\}_{0 = 1}^{2}$ of one-forms on $\scrH$ out of the `old un-primed basis covectors' as follows
	\begin{equation*}
	 \ehat^{\prime i}_{a}~\hat{=}~\ehat^{i}_{b}\tensor{{\pmet'}}{^{b}_{a}}~,
	\end{equation*}
such that
	\begin{equation*}
	 \ehat^{\prime i}_{a}n^{\prime a}~\hat{=}~0~.
	\end{equation*}
However, since these new basis elements still satisfy
	\begin{equation*}
	 \fhat_{i}^{a}\ehat^{\prime j}_{a} = \tensor{\delta}{_{i}^{j}}~,
	\end{equation*}
we can use these `old un-primed basis vectors' in conjunction with the `new primed basis covectors'\footnote{This does not violate the well-known theorem on uniqueness of the dual basis; the bases in a four dimensional sense are $\left\{\{\Dl_{a}\Omega,\,\ehat^{i}_{a}\},\,\{n^{a},\,\fhat_{i}^{a}\}\right\}$ and $\left\{\{\Dl_{a}\Omega,\,\ehat^{\prime i}_{a}\},\,\{n^{\prime a},\,\fhat_{i}^{a}\}\right\}$.}. The tensor $\qmet^{\prime a b}$, being identical with $\qmet^{a b}$ on $\scrH$~\Eqref{q'met^ab:H}, naturally admits an expansion of the form
	\begin{equation*}
	 \qmet^{\prime a b}~\hat{=}~h^{i j}\fhat_{i}^{a}\fhat_{j}^{b}~.
	\end{equation*}
On the other hand, for the tensor $\qmet'_{a b}$ one may easily verify
	\begin{equation*}
	 \qmet'_{a b}~\hat{=}~h_{i j}\ehat^{\prime i}_{a}\ehat^{\prime j}_{b}~.
	\end{equation*}
Clearly,  the metric induced on $\scrH$ by both $\qmet_{a b}$ and $\qmet'_{a b}$ are the~\emph{same} unit hyperboloid metric.
\subsection{Proof of Theorem~\ref{thm:HPVF}}\label{proof:HPVF}
\begin{proof}
Let $(\hat{\man}, \Omega)$ be a unit hyperboloid completion and $X^{a}$ be a vector in a neighborhood $\nhood$ of $\scrH$. Define the scalar $u$ as well as the vectors $U^{a}$ and $V^{a}$ as follows
	\begin{equation*}
	 u~\equiv~\Omega^{-1}X^{a}\Dl_{a}\Omega~, \qquad\qquad U^{a}~\equiv~\frac{\Omega u}{F}n^{a}~, \qquad\qquad V^{a}~\equiv~X^{a} - U^{a}~.
	\end{equation*}
Clearly, $V^{a}$ is tangential to the $\Omega =$ constant hypersurfaces. 
If $X^{a}$ admits a finite limit to $\scrH$ then the scalar $u$ (and hence also the vector $U^{a}$) as well as the vector $V^{a}$ admit finite limits to $\scrH$ (and~\emph{vice versa}). The latter also means that the one-form $V_{a}~\equiv~\qmet_{a b}V^{b}$ admits a finite limit to $\scrH$. Following Definition~\ref{def:HPVF}, we will henceforth assume that $X^{a}$ admits a finite limit to $\scrH$.

Taking the Lie derivative of $\Omega$ along $X^{a}$ yields
	\begin{equation*}
	 \Lie_{X}\Omega = u\Omega~.
	\end{equation*}
Furthermore, since the mass aspect $\mu$ and its four-gradient $\Dl_{a}\mu$ (equivalently the scalar $(\dl\mu/\dl\Omega)$ and the one-form $D_{a}\mu$) both admit finite limits to $\scrH$ by assumption (motivated by expected properties of physical solutions), so does $\Lie_{X}\mu$, based on the following formula as well as our assumptions on the vector $X^{a}$
	\begin{equation*}
	 \Lie_{X}\mu = u\Omega\frac{\dl\mu}{\dl\Omega} + V^{c}D_{c}\mu~.
	\end{equation*}
Hence, from the expression~\Eqref{def:mu} relating $\mu$ to $F$, we have
	\begin{equation*}
	 \Lie_{X}F = \Omega(\Lie_{X}\mu + u\mu)~.
	\end{equation*}
The above relation shows that $\Lie_{X}F$ actually vanishes on $\scrH$.

Cartan's formula applied to the Lie derivative of $\Omega$ along $X^{a}$ next yields
	\begin{equation*}
	 \Lie_{X}\Dl_{a}\Omega = \ss\Dl_{a}\Omega + \Omega D_{a}u~, \qquad\qquad \ss~\equiv~u + \Omega\frac{\dl u}{\dl\Omega}~.
	\end{equation*}
Some straight-forward computations also yield the following formulae for the Lie derivatives of the `kinematic' fields $n^{a}$ and $\qmet^{a b}$, along $X^{a}$
	\begin{eqnarray*}
	 & \Lie_{X}n^{a} = \left[\frac{\Lie_{X}F}{F} - \ss\right]n^{a} + \bn^{a}~, \\
	 & \Lie_{X}\qmet^{a b} = -\frac{\Omega(n^{a}D^{b}u + n^{b}D^{a}u)}{F} - \frac{2u\Omega}{F}\CMcal{K}^{a b} - D^{a}V^{b} - D^{b}V^{a}~,
	\end{eqnarray*}
where
	\begin{eqnarray*}
	 & \bn^{a}~\equiv~\tensor{\pmet}{^{a}_{c}}(\Lie_{V}n^{c})~, \qquad \bn_{a}~\equiv~\qmet_{a b}\bn^{b}~, \\
	 & \CMcal{K}_{a b}~\equiv~\frac{1}{2}\Lie_{n}\qmet_{a b}~, \qquad\quad \CMcal{K}^{a b}~\equiv~\qmet^{a c}\qmet^{b d}\CMcal{K}_{c d}~.
	\end{eqnarray*}
Note that both $\bn^{a}$ and $\CMcal{K}_{a b}$ are `purely tangential' in the sense $\bn^{a}\Dl_{a}\Omega = 0$ and $\CMcal{K}_{a b}n^{b} = 0$. Also, $\CMcal{K}_{a b}$ admits a finite limits to $\scrH$ by the requirements of Definitions~\ref{def:AR:asymp}-~\ref{def:AR:AM}. Based on the results presented above, we may infer that the tensors $\Omega^{-1}\Lie_{X}\Omega$, $\Lie_{X}\Dl_{a}\Omega$, $\Lie_{X}n^{a}$ and $\Lie_{X}\qmet^{a b}$ admit finite asymptotic limits~\emph{if and only if} the scalar $u$, the one-form $\Omega\Dl_{a}u$, the vectors $V^{a}$ and $\bn^{a}$ and the $(0, 2)$-tensor\footnote{It may appear that we should only require a finite asymptotic limit of the symmetric $(0, 2)$-tensor $D_{(a}V_{b)}$ to exist. However, even if we do that, it would mean $V_{a}$ is at least $C^{1}$ on $\scrH$ and hence $D_{[a}V_{b]}$ should also exist (especially given the topology of $\scrH$). Hence, requiring both $V_{a}$ and $D_{(a}V_{b)}$ to exist is equivalent to claiming $D_{a}V_{b}$ to exist as well. Note that by this assumption, one also has finite asymptotic limits of the tensors $D_{a}V^{b} \equiv \qmet^{b c}D_{a}V_{c}$ as well as $D^{a}V^{b} \equiv \qmet^{a c}\qmet^{b d}D_{c}V_{d}$.} $D_{a}V_{b}$ admit finite limits to $\scrH$. Following Definition~\ref{def:HPVF}, we will henceforth assume that all the necessary properties mentioned above hold. Hence, $X^{a}$ is $\scrH$-preserving with respect to the completion $(\hat{\man}, \Omega)$.

Furthermore, the identity
	\begin{equation*}
	 \Lie_{X}\qmet_{a b} = -F^{-1}(\bn_{a}\Dl_{b}\Omega + \bn_{b}\Dl_{a}\Omega) - \qmet_{a c}\qmet_{b d}(\Lie_{X}\qmet^{c d})~,
	\end{equation*}
demonstrates that the tensor $\Lie_{X}\qmet_{a b}$ can be fully determined from the knowledge of $\Lie_{X}n^{a}$ and $\Lie_{X}\qmet^{a b}$. Indeed, either from this observation or from direct computation one obtains
	\begin{equation*}
	 \Lie_{X}\qmet_{a b} = -\frac{\bn_{a}\Dl_{b}\Omega + \bn_{b}\Dl_{a}\Omega}{F} + \frac{2u\Omega}{F}\CMcal{K}_{a b} + D_{a}V_{b} + D_{b}V_{a}~.
	\end{equation*}
This confirms that $\Lie_{X}\qmet_{a b}$ must admit a finite asymptotic limit as long as $\Lie_{X}n^{a}$ and $\Lie_{X}\qmet^{a b}$ do the same. 

Conversely, it is easy to see that if $X^{a}$ admits a finite asymptotic limit, then the analogues of the vectors $U^{a}$ and $V^{a}$ with respect to another completion $(\hat{\man}, \Omega')$ also admit finite asymptotic limits. Indeed, it turns out that the scalar $u' \equiv (\Omega')^{-1}X^{a}\Dl_{a}\Omega'$ satisfies
	\begin{equation*}
	 u'~\hat{=}~u~, \qquad\qquad \lim_{\Omega \to 0}\Omega\Dl_{a}u' = \lim_{\Omega \to 0}\Omega\Dl_{a}u~.
	\end{equation*}
These conditions, amounts to the existence of finite asymptotic limits of the scalar $(\Omega')^{-1}\Lie_{X}\Omega'$ and the one-form $\Lie_{X}\Dl_{a}\Omega'$. Moreover, the conditions in~\Eqref{Theta:sing} and~\Eqref{Theta'':sing} imply the following asymptotic limits
	\begin{eqnarray*}
	 & \Lie_{X}\Theta~\hat{=}~0~, \\
	 & \Lie_{X}\alpha~\hat{=}~0~, \\
	 & \Lie_{X}\beta~\hat{=}~0~.
	\end{eqnarray*}
These results, applied to the transformation rules~\Eqref{expr:n'^a} and~\Eqref{expr:q'met^ab} yield
	\begin{eqnarray*}
	 & \Lie_{X}n^{\prime a}~\hat{=}~\Lie_{X}n^{a} + (\Lie_{X}\qmet^{a b})D_{b}\omegaf + D^{a}(X^{c}D_{c}\omegaf)~, \\
	 & \Lie_{X}\qmet^{\prime a b}~\hat{=}~\Lie_{X}\qmet^{a b}~.  
	\end{eqnarray*}
Clearly, all the necessary limits exist with respect to the completion $(\hat{\man}, \Omega')$ as well (finiteness of the limit of $\Lie_{X}\qmet'_{a b}$ is guaranteed just like the way that of $\Lie_{X}\qmet'_{a b}$ is guaranteed). Hence, if a vector field is $\scrH$-preserving with respect to one unit hyperboloid completion, then it is so with respect to all unit hyperboloid completions.
\end{proof}
\subsection{Proof of Theorem~\ref{thm:w}}\label{proof:w}
\begin{proof}
Let $\xi^{a}$ be a $\scrH$-preserving vector field according to the requirements of Theorem~\ref{thm:HPVF}. Then, the following hold
	\begin{eqnarray*}
	 & \Lie_{\xi}\Omega = u\Omega~, \\
	 & \Lie_{\xi}\Dl_{a}\Omega = \ss\Dl_{a}\Omega + \Omega D_{a}u~, \qquad\qquad \ss~\equiv~u + \Omega\frac{\dl u}{\dl\Omega}~,
	\end{eqnarray*}
as observed in the proof of Theorem~\ref{thm:HPVF}, where the function $u$ admits a finite asymptotic limit and so does $\Omega(\dl u/\dl\Omega)$. Next, taking a Lie derivative of both sides of the definition of $n^{a}$~\Eqref{def:n^a} along $\xi^{a}$ yields
	\begin{equation*}
	 \Lie_{\xi}n^{a} = (\ss - 4u)n^{a} + \Omega^{-4}(\Lie_{\xi}\met^{a b})\Dl_{b}\Omega + \Omega^{-1}D^{a}u~.
	\end{equation*}
Now, if $\xi^{a}$ is also an asymptotic symmetry generator according to Definition~\ref{def:ASAM}, then by the requirements~\Eqref{def:ASAM:Lxi_dO:H} and~\Eqref{def:ASAM:AIIPM} we must have
	\begin{eqnarray*}
	 & \lim_{\Omega \to 0}\ss = 0~, \\
	 & \lim_{\Omega \to 0}(\ss - 4u)  = 0~, \\
	 & \lim_{\Omega \to 0}\Omega D_{a}u  = 0~, \\
	 & \lim_{\Omega \to 0}\Omega^{-1}D^{a}u  \neq \infty~.
	\end{eqnarray*}
In particular, the second and the fourth conditions presented above hold because, $\xi^{a}$ being a $\scrH$-preserving vector field, the vector $\Lie_{\xi}n^{a}$ must be finite asymptotically and become tangential to the $\Omega =$ constant hypersurfaces in the same limit. The unique solution to the above conditions is
	\begin{eqnarray*}
	 & \lim_{\Omega \to 0}u = 0~, \\
	 & \lim_{\Omega \to 0}\Omega\frac{\dl u}{\dl\Omega} = 0~, \\
	 & \lim_{\Omega \to 0}D_{a}u = 0~.
	\end{eqnarray*}
Now, define the function
	\begin{equation*}
	 w~\equiv~\frac{u}{\Omega}~.
	\end{equation*}
Clearly, $w$ must be at least $C^{1}$ in $\nhood$ but can diverge on $\scrH$. However, such a divergence must be restricted by the various conditions on the function $u$ derived above. These conditions, when expressed in terms of the function $w$, become
	\begin{eqnarray*}
	 & \lim_{\Omega \to 0}\Omega w = 0~, \\
	 & \lim_{\Omega \to 0}\Omega^{2}\frac{\dl w}{\dl\Omega} = 0~, \\
	 & \lim_{\Omega \to 0}D_{a}w \neq \infty~.
	\end{eqnarray*}
These are precisely the equations presented in equation~\Eqref{w:sing}.
\end{proof}
For the sake of completeness, let us record the following formul{\ae} describing the action of a general asymptotic symmetry generator on the function $\Omega$ and the one-form $\Dl_{a}\Omega$:
	\begin{eqnarray*}
	 & \Lie_{\xi}\Omega = \Omega^{2}w~, \\
	 & \Lie_{\xi}\Dl_{a}\Omega = \left[\Omega^{2}\frac{\dl w}{\dl\Omega} + 2\Omega w\right]\Dl_{a}\Omega + \Omega^{2}D_{a}w~.
	\end{eqnarray*}
\subsection{Proof of Theorem~\ref{thm:ASAM:identify}}\label{proof:ASAM:identify}
The statement of Theorem~\ref{thm:ASAM:identify} suggests an algorithm to determine whether a given vector field is an asymptotic symmetry generator. The following proof essentially fleshes out the details of that algorithm.
\begin{proof}
Given a $\scrH$-preserving vector field $\xi^{a}$, we may recall equation~\Eqref{def:w} to `read off' the scalar function $w$ associated with it. Suppose $w$ satisfies the conditions in~\Eqref{w:sing}. Now construct the vector $\tau^{a} \equiv F^{-1}\Omega^{2}wn^{a} - \Omega D^{a}w$. Then $\tau^{a}$ is a $\scrH$-preserving vector field by construction. Therefore, so is $\psi^{a} \equiv \xi^{a} - \tau^{a}$, being a linear combination of two $\scrH$-preserving vector fields. Furthermore, since $\psi^{a}$ is tangential to the $\Omega =$ constant hypersurfaces by definition, it satisfies the following equations (as consequences of various relevant definitions)
	\begin{equation*}
	 \Lie_{\psi}\Dl_{a}\Omega = 0~, \qquad\qquad\qquad \Lie_{\psi}n^{a} = \Omega^{-4}(\Lie_{\psi}\met^{a b})\Dl_{b}\Omega~,
	\end{equation*}
and both of these equations admit finite limits to $\scrH$ due to the fact that $\psi^{a}$ is $\scrH$-preserving.

Observe, since $\tau^{a}$ satisfies equations~\Eqref{expr:Ltau_O}-~\Eqref{Ltau_qmet^ab:H} (it is, in fact, an asymptotic symmetry generator by construction). Therefore, if we also suppose that $\psi^{a}$ satisfies $\Lie_{\psi}n^{a}~\hat{=}~0$ and $\Lie_{\psi}\qmet^{a b}~\hat{=}~0$, then linearity of limits and Lie derivatives dictate that $\xi^{a}$ satisfies all the criteria of an asymptotic symmetry generator according to Definition~\ref{def:ASAM}.
\end{proof}
In simple terms, the algorithm says that given a $\scrH$-preserving vector field, subtract off the `$\tau$-part' from it. If the left over vector field is an asymptotic Killing vector of the unit hyperboloid metric on $\scrH$, then the original vector field is an asymptotic symmetry generator.
\subsection{Statement and proof of Theorem~\ref{thm:linearity}}\label{append:linearity}
	\begin{theorem}[Linearity of asymptotic symmetry generators of AM spacetimes]\label{thm:linearity}
Let $\xi_{1}^{a}$ and $\xi_{2}^{a}$ be asymptotic symmetry generators of an asymptotically Minkowskian spacetime with respect to a given unit hyperboloid completion $(\hat{\man}, \Omega)$, as per Definition~\ref{def:ASAM}. Then so is their linear combination
	 \begin{equation*}
	  \xi^{a}~\equiv~c_{1}\xi_{1}^{a} + c_{2}\xi_{2}^{a}~,
	 \end{equation*}
for arbitrary constants $c_{1}$ and $c_{2}$. Consequently, in terms of the functions
	 \begin{equation*}
	  w_{1}~\equiv~~\Omega^{-2}\Lie_{\xi_{1}}\Omega~, \qquad w_{2}~\equiv~~\Omega^{-2}\Lie_{\xi_{2}}\Omega~,
	\end{equation*}
both of which satisfies conditions~\Eqref{w:sing} by assumption, if we define the vector fields
	 \begin{eqnarray}
	  \tau_{1}^{a}~&\equiv&\frac{\Omega^{2}w_{1}}{F}n^{a} - \Omega D^{a}w_{1}~, \qquad\qquad \psi_{1}^{a}~\equiv~\xi_{1}^{a} - \tau_{1}^{a}~, \\
	  \tau_{2}^{a}~&\equiv&~\frac{\Omega^{2}w_{2}}{F}n^{a} - \Omega D^{a}w_{2}~, \qquad\qquad \psi_{2}^{a}~\equiv~\xi_{2}^{a} - \tau_{2}^{a}~,
	 \end{eqnarray}
then $\xi^{a}$ admits a decomposition of the form
	 \begin{equation*}
	  \xi^{a} = \tau^{a} + \psi^{a}~,
	 \end{equation*}
where
	 \begin{equation*}
	  \tau^{a}~\equiv~c_{1}\tau_{1}^{a} + c_{2}\tau_{2}^{a}~, \qquad \psi^{a}~\equiv~c_{1}\psi_{1}^{a} + c_{2}\psi_{2}^{a}~.
	 \end{equation*}
$\tau^{a}$ vanishes asymptotically while $\psi^{a}$ as well as $\xi^{a}$ approach a Killing vector of the unit hyperboloid metric on $\scrH$ in the same limit.
	\end{theorem}
\begin{proof}
As already hinted in the main text, the theorem is a simple consequence of linearity of limits and Lie derivatives.

To begin with, since $\xi_{1}^{a}$ and $\xi_{2}^{a}$ are asymptotic symmetry generators by assumption, and hence $\scrH$-preserving, so is their linear combination $\xi^{a}$.

Next, given a unit hyperboloid completion $(\hat{\man} ,\Omega)$, let the function $w$ be defined as
	\begin{equation*}
	 w~\equiv~\Omega^{-2}\Lie_{\xi}\Omega~.
	\end{equation*}
Linearity of Lie derivatives then indicate
	\begin{eqnarray*}
	 w &=& \Omega^{-2}(c_{1}\Lie_{\xi_{1}}\Omega + c_{2}\Lie_{\xi_{2}}\Omega) \\
	   &=& c_{1}w_{1} + c_{2}w_{2}~.
	\end{eqnarray*}
Clearly $w$ satisfies the conditions~\Eqref{w:sing} since $w_{1}$ and $w_{2}$ do the same by assumption.

Now, let $\mathbf{T}$ be a placeholder for any of the `kinematic' fields associated with the completion $(\hat{\man}, \Omega)$. Hence $\mathbf{T}$ admits a finite limit to $\scrH$ and so do $\Lie_{\xi}\mathbf{T}$, $\Lie_{\xi_{1}}\mathbf{T}$ and $\Lie_{\xi_{2}}\mathbf{T}$ (since $\xi^{a}$, $\xi_{1}^{a}$ and $\xi_{2}^{a}$ are $\scrH$-preserving). In particular, linearity of limits and Lie derivatives imply
	\begin{equation*}
	 \Lie_{\xi}\mathbf{T}~\hat{=}~c_{1}\Lie_{\xi_{1}}\mathbf{T} + c_{2}\Lie_{\xi_{2}}\mathbf{T}~.
	\end{equation*}
Thus, for appropriate choices of the tensor $\mathbf{T}$, the vector field $\xi^{a}$ satisfies all the conditions of Definition~\ref{def:ASAM}. Hence, it is an asymptotic symmetry generator.

Consequently, by Theorem~\ref{thm:ASAM:props}, one may express $\xi^{a}$ as the sum of vector fields $\tau^{a}$ and $\psi^{a}$, where the former is given by
	\begin{eqnarray*}
	 \tau^{a} &=& \frac{\Omega^{2}w}{F}n^{a} - \Omega D^{a}w \\
	          &=& \frac{\Omega^{2}(c_{1}w_{1} + c_{2}w_{2})}{F}n^{a} - \Omega D^{a}(c_{1}w_{1} + c_{2}w_{2}) \\
	          &=& c_{1}\tau_{1}^{a} + c_{2}\tau_{2}^{a}~,
	\end{eqnarray*}
while the latter, also by Theorem~\ref{thm:ASAM:props}, is given by
	\begin{eqnarray*}
	 \psi^{a} &=& \xi^{a} - \tau^{a} \\
	          &=& (c_{1}\xi_{1}^{a} + c_{2}\xi_{2}^{a}) - (c_{1}\tau_{1}^{a} + c_{2}\tau_{2}^{a}) \\
	          &=& c_{1}\psi_{1}^{a} + c_{2}\psi_{2}^{a}~.
	\end{eqnarray*}
Since $\psi_{1}^{a}$ and $\psi_{2}^{a}$ satisfy the Killing equation asymptotically~\Eqref{KVE:psi:H}, so does their linear combination $\psi^{a}$. Note that $\tau_{1}^{a}$, $\tau_{2}^{a}$, $\psi_{1}^{a}$ and $\psi_{2}^{a}$ are all asymptotic symmetry generators, and so are their linear combinations $\tau^{a}$ and $\psi^{a}$.

Finally, $\tau^{a}$ vanishes in the asymptotic limit by construction, leaving us with $\xi^{a}~\hat{=}~\psi^{a}$. Hence $\psi^{a}$ (as well as $\xi^{a}$) acts as a Killing vector of the unit hyperboloid metric on $\scrH$ asymptotically.
\end{proof}
\subsection{Verifying completion independence}\label{append:comp_indep}
In this appendix, we will briefly discuss how completion independence can be properly formulated and explicitly verified in a given context.

Suppose that a result -- \ie, some mathematical expression satisfying some well-defined property -- holds with respect to a given completion $(\hat{\man}, \Omega)$. Then the result will be completion independent if the said mathematical expression, rewritten in terms of the kinematic fields associated with another completion $(\hat{\man}, \Omega')$ also satisfies the same property.

The general strategy to `translate' a mathematical expression from $(\hat{\man}, \Omega)$ to $(\hat{\man}, \Omega')$ is as follows:
	\begin{itemize}
	 \item Use the transformation rules~\Eqref{def:Omega'},~\Eqref{expr:dO'},~\Eqref{expr:n'^a},~\Eqref{expr:q'met_ab} and~\Eqref{expr:q'met^ab}, along with the expressions for $F$ in terms of the mass aspect~\Eqref{def:mu}, as well as those for $\alpha$~\Eqref{def:alpha} and $\beta$~\Eqref{expr:beta} in terms of $\omega$ and its derivatives. 
	 \item Construct the `primed projector' in terms of the `primed' kinematic fields according to the expression~\Eqref{def:pmet}, and use it to define the `primed projected covariant derivative' $D'_{a}$ compatible with $\qmet'_{a b}$.
	 \item Utilize the observation in footnote~\Eqref{ftnt:Lien=ddO} to convert `$\Omega$-derivatives' into `$\Omega'$-derivatives'.
	\end{itemize}
Finally, the asymptotic limit of any mathematical expression with respect to any completion can be taken by allowing the corresponding completion function to approach the value zero. The above strategy is adequate to establish completion independence of, for example, the results in equations~\Eqref{omega:sing},~\Eqref{Theta:sing},~\Eqref{omega':sing},~\Eqref{Theta:gauge},~\Eqref{Theta'':sing},~\Eqref{w:sing} and~\Eqref{w':sing}, the general expression~\Eqref{expr:Lambda^a} of a gauge generator, as well as the results in various definitions and theorems.

\subsection{A unit hyperboloid completion of the Schwarzschild spacetime}\label{append:Schwarzschild_UHC}
In this appendix, we want to briefly illustrate how to construct a unit hyperboloid completion of the Schwarzschild spacetime which, furthermore, makes it possible to cast the Schwarzschild metric in the standard Beig-Schmidt form~\cite{Beig:1982ifu, Ashtekar:1991vb}\footnote{A similar analysis presented in section 3 of reference~\cite{Ashtekar:1991vb} is inadequate since the metric could not be brought to a Beig-Schmidt form there.}

It is very simple to argue that a spherically symmetric asymptotically flat spacetime can always be expressed in a Beig-Schmidt type coordinate system with the help of the decomposition~\Eqref{expr:met_ab}, as follows
    \begin{equation*}
        \met_{a b} = \frac{\Dl_{a}\Omega\Dl_{b}\Omega}{F\Omega^{4}} - \frac{N^{2}\Dl_{a}\chi\Dl_{b}\chi}{\Omega^{2}} + r^{2}\bar{\gamma}_{a b}~,
    \end{equation*}
where
    \begin{itemize}
     \item the functions $\chi$, $\theta$ and $\varphi$ are, respectively, the standard `hyperbolic' time and spherical polar coordinate functions on a unit hyperboloid;
     \item the rank-$(0, 2)$ tensor $\bar{\gamma}_{a b}$ denotes the standard metric on a unit two-sphere
    \begin{equation*}
        \bar{\gamma}_{a b}~\equiv~\Dl_{a}\theta\Dl_{b}\theta + \sin^{2}\theta\Dl_{a}\varphi\Dl_{b}\varphi~;
    \end{equation*}
     \item the functions $F$, $N$ and $r$ are all independent of the spherical polar coordinates, the first two become unity asymptotically, while the last one is the standard~\emph{areal radial function};
     \item the function $F$ can be cast in a form such that
    \begin{equation*}
     \frac{1}{\sqrt{F}} = 1 - \frac{\muH\Omega}{2}~,
    \end{equation*}
where $\muH$ is the asymptotic limit of the mass aspect function;
     \item the `shift-terms', potentially arising from a $(3 + 1)$-decomposition of the metric with respect to the $\Omega =$ constant hypersurfaces, are all zero.
    \end{itemize}
Consider, now, going from the system of coordinates $\{\Omega, \chi, \theta, \varphi\}$ to $\{t, r, \theta, \varphi\}$, where $t = t(\Omega, \chi)$ and $r = r(\Omega, \chi)$, such that the metric in the latter chart takes the well-known form
    \begin{equation*}
     \met_{a b} = -\left(1 - \frac{\rs}{r}\right)\Dl_{a}t\Dl_{b}t + \left(1 - \frac{\rs}{r}\right)^{-1}\Dl_{a}r\Dl_{b}r + r^{2}\bar{\gamma}_{a b}~.
    \end{equation*}
The Beig-Schmidt requirements listed above then imply
    \begin{eqnarray*}
     \Omega^{4}\left[-\left(1 - \frac{\rs}{r}\right)\left(\frac{\dl t}{\dl\Omega}\right)^{2} + \left(1 - \frac{\rs}{r}\right)^{-1}\left(\frac{\dl r}{\dl\Omega}\right)^{2}\right] - \left(1 - \frac{\muH\Omega}{2}\right)^{2} & = 0~, \\
      \Omega^{3}\left[-\left(1 - \frac{\rs}{r}\right)\left(\frac{\dl t}{\dl\Omega}\right)\left(\frac{\dl t}{\dl\chi}\right) + \left(1 - \frac{\rs}{r}\right)^{-1}\left(\frac{\dl r}{\dl\Omega}\right)\left(\frac{\dl r}{\dl\chi}\right)\right] & = 0~,
    \end{eqnarray*}
via the usual rule of coordinate transformations. The above equations can be solved, \eg, by expressing the functions $t$ and $r$ as power series in $\Omega$ whose coefficients are functions of $\chi$, and solving for each coefficient order by order. In particular, we find
    \begin{eqnarray*}
        t & = \frac{\sinh\chi}{\Omega} + t_{0} + \frac{\rs^{2}\sinh\chi}{2}\left[-\frac{4\log\cosh\chi}{\cosh^{2}\chi} + \frac{1}{2\cosh^{2}\chi} + 1\right]\Omega + \order(\Omega^{2})~, \\
        r & = \frac{\cosh\chi}{\Omega} - 2\rs\log\cosh\chi + \rs^{2}\left[\frac{3}{4\cosh\chi} + \frac{\log\cosh\chi}{\cosh^{3}\chi} - \frac{3}{8\cosh^{3}\chi}\right]\Omega + \order(\Omega^{2})~.
    \end{eqnarray*}
The leading order terms are obviously indicative of asymptotic flatness of the metric, and the constant $t_{0}$ is a relic of time translation symmetry. Furthermore, the asymptotic limit of the mass aspect is given by
    \begin{equation*}
        \muH = \left[\frac{1}{\cosh\chi} - 2\cosh\chi\right]\rs~.
    \end{equation*}
This agrees with the result of reference~\cite{Ashtekar:1991vb}, suggesting that the unit hyperboloid completion proposed in that work is gauge equivalent to the one presented above. Finally, the function $N$ is given by
    \begin{eqnarray*}
        N & = \Omega\left[\left(1 - \frac{\rs}{r}\right)\left(\frac{\dl t}{\dl\chi}\right)^{2} - \left(1 - \frac{\rs}{r}\right)^{-1}\left(\frac{\dl r}{\dl\chi}\right)^{2}\right]^{\frac{1}{2}}~, \\
          & = 1 + \rs\left[\cosh\chi - \frac{3}{2\cosh\chi}\right]\Omega + \order(\Omega^{2})~.
    \end{eqnarray*}
The consistency of the above results in the $\rs \to 0$ limit should be obvious.

While the Beig-Schmidt form of the metric presented above makes the spherical symmetry of the spacetime manifest, it certainly obscures the fact that the metric is static as well. However, one may construct the time translation generating Killing vector $\KVT^{a}$ simply by requiring
    \begin{equation*}
        \KVT^{a}\Dl_{a}t = 1~, \qquad \KVT^{a}\Dl_{a}r = 0~, \qquad \KVT^{a}\Dl_{a}\theta = 0~, \qquad \KVT^{a}\Dl_{a}\varphi = 0~,
    \end{equation*}
which ensures that $\KVT^{a}$ is the sought after Killing vector. In the Beig-Schmidt coordinates, its expression works out to be
    \begin{eqnarray*}
     \KVT^{a} & = \frac{1}{\left[-(\dl r/\dl\chi)(\dl t/\dl\Omega) + (\dl r/\dl\Omega)(\dl t/\dl\chi)\right]}\left[-\left(\frac{\dl r}{\dl\chi}\right)\left(\frac{\dl}{\dl\Omega}\right)^{a} + \left(\frac{\dl r}{\dl\Omega}\right)\left(\frac{\dl}{\dl\chi}\right)^{a}\right] \\
             & = \Omega^{2}\sinh\chi\left(\frac{\dl}{\dl\Omega}\right)^{a} + \Omega\cosh\chi\left(\frac{\dl}{\dl\chi}\right)^{a} + \Lambda^{a}~,
    \end{eqnarray*}
where
    \begin{equation*}
        \Lambda^{a}~\equiv~-2\Omega^{2}\rs\left[\left(\cosh\chi\sinh\chi\Omega + \order(\Omega^{2})\right)\left(\frac{\dl}{\dl\Omega}\right)^{a} + \left(\sinh^{2}\chi + \order(\Omega)\right)\left(\frac{\dl}{\dl\Omega}\right)^{a}\right].
    \end{equation*}
Clearly, $\Lambda^{a}$ is gauge generator and hence the Killing vector $\KVT^{a}$ is gauge equivalent to an asymptotic time translation symmetry generator.

\section*{References}
\bibliographystyle{unsrt}
\bibliography{LogST_BSAR}
\end{document}